\documentclass[useAMS,usenatbib,useasmath]{mnras}
\usepackage{graphicx}
\usepackage{float}
\usepackage{longtable}
\usepackage{times}
\usepackage{amsmath}
\usepackage{amssymb}
\usepackage{bm}
\usepackage[toc,page]{appendix}
\usepackage{pdflscape}
\usepackage{xcolor}
\usepackage{hyperref}
\hypersetup{
	colorlinks = true,
	urlcolor = blue,
	linkcolor = blue,
	citecolor = blue
}

\def\aj{AJ}
\def\pasp{PASP}
\def\apj{ApJ}
\def\apjs{ApJS}
\def\apjl{ApJL}
\def\aap{A\&A}
\def\araa{ARA\&A}

\def\mnras{MNRAS}
\def\nat{Nature}
\def\iaucirc{IAU~Circ.}

\newcommand{\Mch}{\hbox{$M_{\text{Ch}}$}}

\newcommand{\Lsun}{\hbox{L$_\odot$}}
\newcommand{\Msun}{\hbox{M$_\odot$}}

\newcommand{\kms}{\hbox{km$\,$s$^{-1}$}}

\newcommand{\elem}[2][default]{$^{#1}{\rm #2}$}

\def\lesssim{\mathrel{\hbox{\rlap{\hbox{\lower4pt\hbox{$\sim$}}}\hbox{$<$}}}}
\def\gtrsim{\mathrel{\hbox{\rlap{\hbox{\lower4pt\hbox{$\sim$}}}\hbox{$>$}}}}
\def\gray{$\gamma$-ray}
\def\grays{$\gamma$-rays}
\def\isoni{$^{56}{\rm Ni}$}

\newcommand{\caii}{\mbox{Ca~{\sc ii}}}
\newcommand{\ariii}{\mbox{Ar~{\sc iii}}}

\newcommand{\oi}{\mbox{O~{\sc i}}}

\newcommand{\skii}{\mbox{Si~{\sc ii}}}
\newcommand{\skiii}{\mbox{Si~{\sc iii}}}
\newcommand{\tkii}{\mbox{Ti~{\sc ii}}}

\newcommand{\feii}{\mbox{Fe~{\sc ii}}}
\newcommand{\feiii}{\mbox{Fe~{\sc iii}}}
\newcommand{\mgii}{\mbox{Mg~{\sc ii}}}
\newcommand{\coii}{\mbox{Co~{\sc ii}}}
\newcommand{\coiii}{\mbox{Co~{\sc iii}}}
\newcommand{\crii}{\mbox{Cr~{\sc ii}}}
\newcommand{\nkii}{\mbox{Ni~{\sc ii}}}
\newcommand{\nkiii}{\mbox{Ni~{\sc iii}}}
\newcommand{\siii}{\mbox{S~{\sc iii}}}
\newcommand{\sii}{\mbox{S~{\sc ii}}}
\newcommand{\lb}{$\lambda$}
\newcommand{\delM}{$\Delta M_{15}(B)$}
\newcommand{\delm}{$\Delta m_{15}(B)$}
\newcommand{\cmfgen}{\sc cmfgen}

\def\fig{Fig.}
\def\figs{Figs.}

\title[SNe Ia Ejecta Mass Diagnostics]
{Ejecta Mass Diagnostics of Type Ia Supernovae}
\author[Kevin D. Wilk, D. John Hillier, Luc Dessart]
{Kevin D. Wilk,$^{1}$\thanks{E-mail: kdw25@pitt.edu} D. John Hillier,$^1$\thanks{E-mail: hillier@pitt.edu}  Luc Dessart$^2$ \\
$^1$ Department of Physics and Astronomy \& Pittsburgh Particle physics, Astrophysics, and Cosmology Center (PITT PACC), University of Pittsburgh,  \\ Pittsburgh, PA 15260, USA \\
$^2$ Unidad Mixta Internacional Franco-Chilena de Astronom\'ia (CNRS UMI 3386), 
Departamento de Astronom\'ia, Universidad de Chile, \\
Camino El Observatorio 1515, Las Condes, Santiago, Chile}
\begin{document}

\date{Accepted . Received }

\pagerange{\pageref{firstpage}--\pageref{lastpage}} \pubyear{2016}

\maketitle

\label{firstpage}

\begin{abstract}
We present one-dimensional non-local thermodynamic equilibrium time-dependent radiative transfer simulations (using {\cmfgen}) of two sub-Chandrasekhar (sub-\Mch), one \Mch\ and one super-\Mch\ Type Ia SN ejecta models. Three originate from \Mch\ delayed detonation models, and the fourth is a sub-\Mch\ detonation model.  Ejecta masses are 1.02, 1.04, 1.40, and 1.70 \Msun, and all models have 0.62 \Msun\ of \elem[56]{Ni}. Sub-\Mch\ model light curves evolve faster, reaching bolometric maximum 2--3 days earlier and having 3--4 days shorter bolometric half light widths. The models vary by $\sim$12 per cent at maximum bolometric luminosity and by 0.17 mag in $B_{\rm max}$. While $\Delta M_{15}(B)$ increases with ejecta mass it only varies by $\sim$5 per cent around 1 mag. Sub-\Mch\ models are 0.25 mag bluer in $B-R$ at $B_{\rm max}$. Optical spectra share many similarities, but lower mass models exhibit less UV line blanketing during the photospheric phase. At nebular times, significant NIR spectroscopic differences are seen. In particular, emission lines of the \caii\ NIR triplet; [\siii] \lb\lb9068,9530; [\caii] \lb\lb7291,7324; [\ariii] \lb\lb7135,7751; and [\nkii] 1.939 $\mu$m are stronger in higher mass models. The [\nkii] 1.939 $\mu$m line is absent in the sub-\Mch\ detonation model, and provides a valuable potential tool to distinguish sub-\Mch\ explosions from \Mch\ explosions. In general, the nebular phase models are too highly ionized. We attribute this to the neglect of clumping and/or the distribution of intermediate mass and iron group elements. The two sub-\Mch\ models, while exploded by different mechanisms, can be distinguished in the $J$ and $H$ bands at late times (e.g., $+200$ days).
\end{abstract}

\section{Introduction}

Type Ia supernovae (SNe Ia) are thought to be thermonuclear runaway explosions of carbon-oxygen (C/O) white dwarfs (WDs) \citep{Hoyle1960}, but the formation channel of the progenitor remains uncertain. Determining the progenitor channel(s) of SNe Ia, and their diversity, is of crucial importance since it will improve their use as probes of cosmological parameters \citep{Riess1998,Perlmutter1999}. Further, understanding the progenitor channel allows us to probe stellar evolution of binary systems prior to the SN occurring, and hence place constraints on both binary synthesis models and binary evolution.

There are two main progenitor channels invoked to explain the origin of SNe Ia -- the single degenerate (SD) channel and double degenerate channel. In the ``classic" SD channel, the WD accretes matter (usually hydrogen and/or helium) from a non-degenerate star due to binary interaction via Roche lobe overflow.
However, mass transfer in the SD channel is not limited to Roche lobe overflow --  it can also be mediated by wind mass loss as, for example, in symbiotic systems \citep[see][for a review]{Maoz2014}. In the SD channel matter accumulates on the white dwarf, where stable burning can occur, until the star approaches the Chandrasekhar mass (\Mch$\approx$1.4 $M_\odot$) and explodes leading to a SN Ia \citep{WhelanIben1973,Nomoto1982a}.

How the WD explodes has long been a matter of study within the astrophysics community. The different explosion mechanisms include pure detonation \citep{Arnett1969,Sim_2010}, pure deflagration \citep{Nomoto_1976}, delayed detonation and pulsating delayed detonation \citep{Khokhlov1991a,Khokhlov1991b, Gamezo2005,Livne2005,Ropke2007,Jackson2010,Seitenzahl2013}, and gravitationally-confined detonation (GCD) \citep{Plewa2004}. More recent research on the GCD, including 3D full star simulations, was presented by \cite{Seitenzahl2016}. For sub-\Mch\ WD masses, a double detonation scenario has been explored \citep{Woosley1994,Livne1995, Fink2007,Fink2010}. 

One of the most successful theories at reproducing the properties of standard SNe Ia is the delayed-detonation transition (DDT) model where the WD first undergoes a subsonic deflagration phase. The resulting expansion of the WD creates lower density fuel, which is necessary for the production of intermediate mass elements (IMEs). At a particular density, the burning switches from subsonic to supersonic \citep{Khokhlov1991a,Khokhlov1991b}. This scenario is required to recreate the observed chemical stratification. Pure detonation models produce too many iron group elements (IGEs) such as iron and nickel but not enough IMEs at high velocities to reproduce SN Ia spectra. On the other hand, deflagration models produce abundant amounts of IMEs but not enough \elem[56]{Ni} to power the SN luminosity. This mechanism is a possible channel for under-luminous SNe Ia \citep{Travaglio2004,Phillips2007,Jordan2012,Kromer2013}.
One dimensional (1D) DDT models have been  very successful in reproducing the observed properties of SNe Ia. However, these models describe the turbulent flame propagation only in a parametrized way. Three dimensional (3D) models better capture the fluid instabilities and mixing. This lead to different ejecta structures that do not reproduce observables as well as 1D models \citep[see e.g.][]{Seitenzahl2013,Sim2013MNRAS}.

The double degenerate scenario corresponds to the merger of two WDs through loss of angular momentum by gravitational wave emission. It is not understood how the merger triggers the explosion, although it is thought that during the merger the smaller mass WD donates matter to the more massive WD. Recently, modeling by \cite{Sato2015} has shown that, depending on the masses of the two WDs, explosions can occur either during the merger phase, provided both WDs are within a mass range between 0.9--1.1 M$_\odot$, or within the merger remnant phase when the more massive object reaches \Mch. From their models, the authors estimate that 9 per cent of galactic SNe Ia can be attributed to mergers. One problem in this scenario is that rapid mass accretion  of carbon and oxygen leads to an off-centre carbon ignition and subsequently an O/Ne/Mg WD \citep{Saio_Nomoto1985}. Accretion induced collapse creates additional problems for high accretion rates, leading to the formation of a neutron star instead of a SN Ia. For an extensive review about progenitors of SNe Ia, see \cite{Maoz2014}.

When almost the entire C/O WD has burnt, releasing $\sim$$10^{51}$ ergs (more energy than the gravitational binding energy of a \Mch\ WD), the energy liberated unbinds the WD, producing an ejecta with no remnant. Ejecta velocities of the order of $\sim$10\,000\,\kms\ are produced, and within minutes, the ejecta reaches a homologous coasting phase (radiation pressure from \elem[56]{Ni} decay produces second order effects).  
At early times the ejecta is radiation dominated and heated by the decay of \elem[56]{Ni} (\elem[56]{Co} for post-maximum evolution).  It is because of this main decay chain of \elem[56]{Ni}$\rightarrow$\elem[56]{Co}$\rightarrow$\elem[56]{Fe}, with roughly 1.7 MeV of energy release per decay for \elem[56]{Ni} and 3.7 MeV per decay for \elem[56]{Co}, that these objects are so luminous. However, \cite{Dessart2014a} show the importance of heating the outer ($\ge$20\,000 \kms) ejecta from other decay channels. 

Early work by \cite{Stritzinger2006a} suggested sub-\Mch\ WDs as progenitors for some SNe Ia based on comparisons of $UVOIR$ light curves of 16 SNe with analytical models of nuclear decay luminosities and energy deposition. \cite{Scalzo2014a,Scalzo2014b}, \cite{Sim_2010,Sim2013}, and \cite{Blondin2017} have shown that there is both observational and theoretical evidence for sub-\Mch\ explosions. From photometric model fitting, \cite{Scalzo2014a,Scalzo2014b} shows that the mass distribution for SNe Ia includes both sub-\Mch\ and super-\Mch\ events. \cite{Scalzo2014b} argue that 25--50 per cent of SN Ia events deviate from \Mch\ events, with most of these occurring as sub-\Mch\ explosions. Therefore, for given measured \elem[56]{Ni} masses, one can compare SNe Ia to better understand how ejecta mass affects light curves and spectral evolution. 

\cite{Sim_2010,Sim2013} found good agreement with photometric observations at maximum and reproduced IME features in synthetic spectra at maximum with 1D pure detonations of sub-\Mch\ explosions. However, these models decline too rapidly post maximum. This idealized approach ignores the influence of any accreted helium layer. Previous works \citep{Woosley1994,Livne1995,HoeflichKhokhlov1996,Hoeflich1996,Nugent1997,Kromer2010} found difficulties reproducing the observed light curves, colours and spectral evolution after treating the burnt helium layer, which synthesizes a significant amount of \elem[56]{Ni}. 

\cite{Woosley2007} explored a grid of SN Ia models coming from 1.38 \Msun\ WDs with varying amounts of mixing and \elem[56]{Ni}, finding models of similar \elem[56]{Ni} to have large variations of decline rates (\delM\ -- which is the change in $B$-band magnitude 15 days after $B$-band maximum \citep{Phillips1993}) and anti width-luminosity relationships. \cite{Woosley2011} computed hydrodynamic and radiative transfer simulations of sub-\Mch\ mass models for helium-accreting WDs, but the authors only found reasonable agreement with spectra and light curves of common SNe Ia for the most massive white dwarfs with the smallest helium layers they considered. \cite{Blondin2017} looks at broadening our understanding of the width-luminosity relation (WLR) using pure detonations of sub-\Mch\ WDs. Their work shows promising agreement with observations of faint SNe Ia, confirming the need for two WD populations to explain the full behavior of the WLR seen at high and low SN Ia brightnesses.

Observations have shown high-velocity features (HVFs) in early-time spectra of SNe Ia. These features have been used to constrain the explosion scenario \citep{Mazzali2005a,Mazzali2005b,Tanaka2006, Blondin2013,Childress2013,Childress2014,Silverman2015,Zhao2015,Pan2015}. Studying the formation of these features as well as their correlation with galaxy environment can improve our understanding of Type Ia progenitors. We can gain insights, for example, into the density and temperature structure of the outer layers. The latter will reveal itself through changes in ionization and hence the strength of spectral features.
 
In this paper we study two sub-\Mch\ models ($\sim$1 \Msun), a 
\Mch\ model ($\sim$1.4 \Msun), and a super-\Mch\ model ($\sim$1.7 \Msun), all with the same \elem[56]{Ni} mass by design,  to determine the effects of ejecta mass. 
One model is a standard DDT model, two models are standard DDT models whose density have been scaled  to give the desired ejecta mass, while the final model arises from a pure detonation in a sub-\Mch\ model.
 The original DDT models are also somewhat artificial. For example, the onset of detonation in a DDT model is a free parameter. Further, it is impossible in current models to resolve and adequately model the thermonuclear flame (see \cite{Ciaraldi-Schoolmann2013} and references therein). We consider evolution over more than two hundred days in time -- from $\sim$1 day after the explosion until approximately 220 days after the explosion, and look for diagnostics of ejecta mass ($M_{\rm ej}$) for our \elem[56]{Ni} mass that can be used to distinguish between the different models. Since the precise explosion mechanism is uncertain, the two sub-\Mch\ models were exploded using different assumptions. Spectra and light curves were computed using non-LTE and time-dependent radiative transfer. Looking for SN Ia diagnostics, we wanted to remove sensitivity of \elem[56]{Ni} while focusing on ejecta mass to uncover insights in SN Ia evolution, both spectroscopically and photometrically.

The paper is organized as follows. In Section \ref{section_technique}, we discuss the techniques used and initial ejecta properties. We compare the bolometric light curves as well as synthetic photometric light curves in Section \ref{section_lc}. We discuss the spectral evolution, highlighting the strong spectral differences, in Section \ref{section_spec_evo}. In Section \ref{section_sub1_vs_sub2}, we further highlight the distinctions between our two sub-\Mch\ models. In Section \ref{section_HVF} we comment on the lack of high velocity features. Section \ref{section_data_compare} shows spectral comparison to a few SNe Ia that are close in measured $M_{\rm ej}$ and \elem[56]{Ni} mass. In Section \ref{section_problems} we discuss shortcomings of our models in reproducing spectral characteristics of SNe Ia beyond 20 days post-maximum. Section \ref{Conclusion} summarizes our results and conclusions.

\section{Technique}%%%%%%%%%%%%%%%%%%%%%%%%%%%%%%%%%%%%%%%%%%%%%%%%%
\label{section_technique}
To determine spectral and light curve diagnostics we utilize hydrodynamical models that have been evolved from explosion until 0.75 days. Homologous expansion of the ejecta is well established at 100 seconds, and it is assumed to strictly hold at all times thereafter. We allow for non-local thermodynamic equilibrium (non-LTE) and solve the spherically symmetric, time-dependent, relativistic radiative transfer equation in order to produce emergent synthetic spectra from which synthetic light curves (LCs) can be produced.

\subsection{Ejecta and Radiative Transfer Modeling} 
Our models correspond to scaled ejecta of model DDC0 (density scaled by 0.73, model SUB2) and DDC15 (density scaled by 1.22, model SUP), complemented with models DDC10 (no scaling applied, model CHAN) -- see \cite{Blondin2013} -- and SCH5p5 (density scaled by 0.98, model SUB1) -- see \cite{Blondin2017}. This density scaling is applied to produce ejecta with the same \elem[56]{Ni} mass initially (0.62 \Msun), but differing in ejecta mass so that they lie below, at, and above the \Mch. This scaling, applied exclusively to the density at 10 seconds after explosion, is obviously artificial. An advantage of this scaling method is that, while the mass varies, the chemical stratification does not. This ensures the models retain their fundamental characteristic of all SN Ia ejecta. We do not compute the combustion nor make any claim that a flame would behave in the way adopted for the corresponding WD mass (i.e., deflagration followed by detonation in the DDC models; pure detonation in the SCH model). Even in the original (unscaled) model, the treatment of combustion is largely imposed rather than computed from first principles. The purpose of the study is to investigate the signatures sensitive to variations in ejecta mass, keeping the \elem[56]{Ni} mass the same between all models in order to retain only one variable quantity. Model masses, kinetic energies, and important species masses are summarized in Table~\ref{model_info_abund}.  

The use of scaled DDT models for SUB2 and SUP is problematical since the explosion properties will depend on the mass of the WD. For example, differences in the initial central densities before  the explosion will likely lead to different abundance profiles within the ejecta. In particular, sub-\Mch\ WDs, unlike their \Mch\ counterpart with stable IGEs in their inner ejecta, do not exhibit a `$^{56}$Ni hole' which is a low \elem[56]{Ni} abundance inside an expansion velocity of 2500 \kms. However, a comparison of SUB2 with SUB1 will allow us to test the sensitivity of the results to the adopted explosion model. The explosion mechanism for ejecta with super-Chandrasekhar masses is extremely uncertain, and any adopted model will have limitations.

The radiative transfer models have been computed using {\cmfgen} \citep{Hillier1998,Hillier2012,Dessart2014a,Dessart2014b}, which solves the spherically symmetric, non-local thermodynamic equilibrium (non-LTE), time-dependent, relativistic radiative transfer equation in the co-moving frame. To advance in time, we used a 10 per cent time step for each model starting from 0.75 days until $\sim$100 days, after which we used a time step of 10 days. At early times ($\lesssim$7 days) during the time sequence, we assumed that $\gamma$-ray photons created from radioactive decays are locally deposited. Otherwise, we approximate the $\gamma$-ray deposition as grey, adopting the procedure from \cite{Swartz1995} and a $\kappa_\gamma=0.06Y_e$ cm$^2$ g$^{-1}$. The kinetic energy of decay positrons is locally deposited at all epochs. {\cmfgen} currently treats both one- and two-step decay chains for calculating non-thermal heating. 

\begin{figure} 
\centering
\includegraphics[scale=0.8]{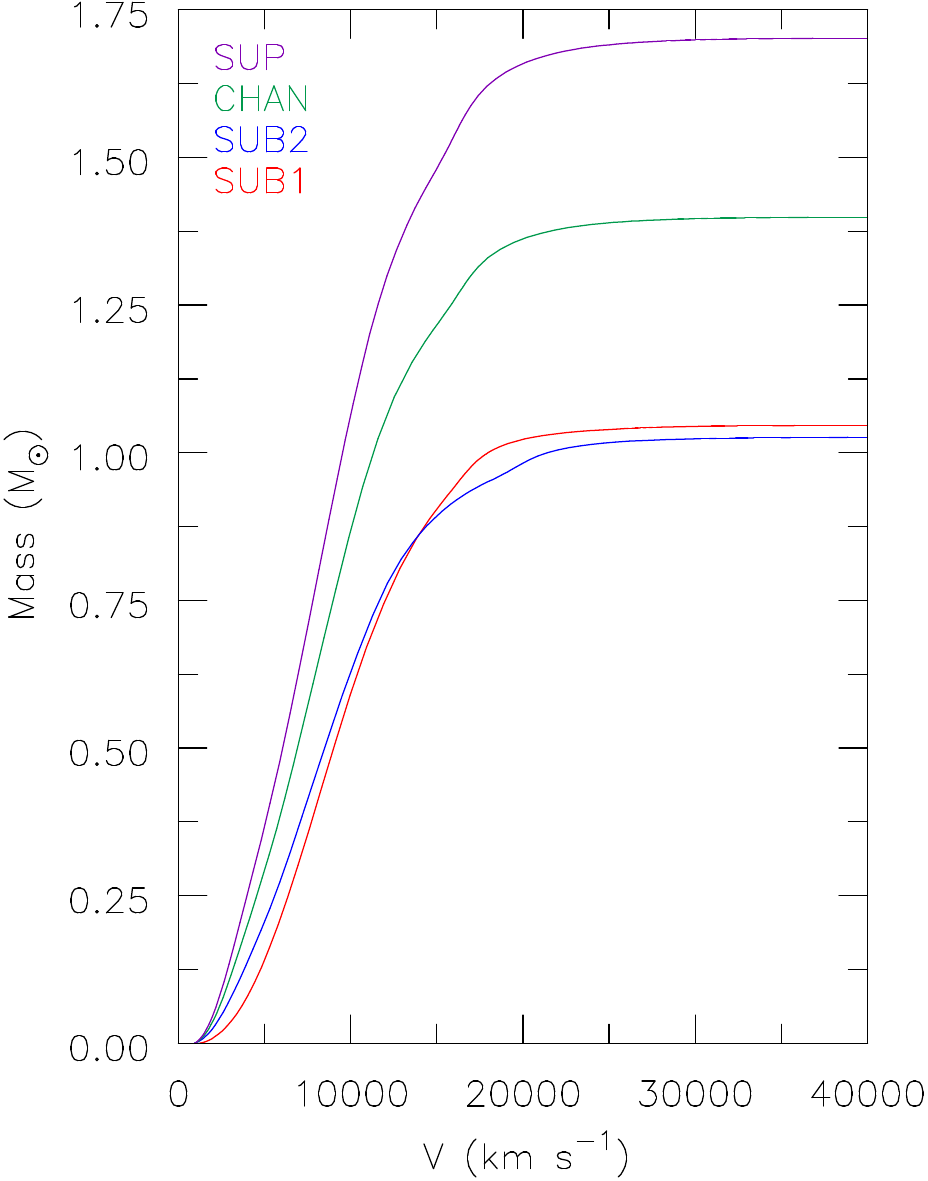}
\caption[]{Illustration of the outward cumulative mass as a function of velocity. The cumulative mass begins to flatten off around 25\,000 \kms\ for all models, the velocity at which the density begins to decrease rapidly. Less than 1 per cent of the total mass is exterior to this velocity.}
\label{mass_vs_v}
\end{figure}

\begin{table*}
\begin{minipage}[t]{\linewidth}
\centering
 \begin{tabular}{| l | c | c | c | c | c | c | c | c | c | c |} 
 \hline
 Model & Mass & $E_{\rm Kin}$ & $\upsilon_{\rm rms}$ & C & O & Na & Mg & Si & S & Ar \\  [0.5ex]
  & (\Msun) & ($10^{51}$ ergs) & (\kms) & (\Msun) & (\Msun) & (\Msun) & (\Msun) & (\Msun) & (\Msun) & (\Msun) \\[0.5ex] 
 \hline\hline
 SUB1 & 1.04 & 1.22 & 10821 & 3.295$(-3)$ & 5.196$(-2)$ & 1.306$(-5)$ & 2.441$(-3)$ & 1.553$(-1)$ & 1.022$(-1)$ & 2.248$(-2)$ \\
 SUB2 & 1.02 & 1.17 & 10695 & 8.803$(-4)$ & 3.906$(-2)$ & 4.466(-6) & 2.615$(-3)$ & 1.185$(-1)$ & 7.488$(-2)$ & 1.654$(-2)$ \\ 
 CHAN & 1.40 & 1.51 & 10415 & 2.419$(-3)$ & 9.595$(-2)$ & 1.244$(-5)$ & 9.557$(-3)$ & 2.551$(-1)$ & 1.661$(-1)$ & 3.693$(-2)$ \\
 SUP  & 1.70 & 1.81 & 10351 & 3.333$(-3)$ & 1.293$(-1)$ & 2.044$(-5)$ & 1.387$(-2)$ & 3.805$(-1)$ & 2.443$(-1)$ & 5.223$(-2)$ \\
 \hline
 Model & Ca & Ti & Cr & Fe & Co & \elem[58]{Ni} + \elem[60]{Ni} & \elem[56]{Ni}  \\ [0.5ex]
& (\Msun) & (\Msun) & (\Msun) & (\Msun) & (\Msun) & (\Msun) & (\Msun)  \\ [0.5ex] 
 \hline\hline
 SUB1 & 2.361$(-2)$ & 2.665$(-5)$ & 1.030$(-2)$ & 2.226$(-2)$ & 5.526$(-2)$ & 1.1$(-2)$ & 5.684$(-1)$  \\
 SUB2 & 1.822$(-2)$ & 1.083$(-4)$ & 1.516$(-2)$ & 6.654$(-2)$ & 5.680$(-2)$ & 2.6$(-2)$ & 5.710$(-1)$  \\ 
 CHAN & 4.120$(-2)$ & 1.486$(-4)$ & 2.689$(-2)$ & 1.020$(-1)$ & 5.713$(-2)$ & 2.5$(-2)$ & 5.708$(-1)$  \\
 SUP  & 5.562$(-2)$ & 1.828$(-4)$ & 2.977$(-2)$ & 1.224$(-1)$ & 5.777$(-2)$ & 3.0$(-2)$ & 5.707$(-1)$  \\
 \hline
 \end{tabular}
\caption[]{Model summary. Listed are the masses in \Msun, ejecta kinetic energies in $10^{51}$ ergs, and the root mean square of the velocity in \kms. Also listed is model mass information  in \Msun\ at 0.75 days post explosion. The parentheses (\#) correspond to $\times10^{\#}$}
\label{model_info_abund}
\end{minipage}
\end{table*}

\subsection{Ejecta Conditions for Radiative Transfer}
Table \ref{model_info_abund} shows the yields for the most abundant species in our ejecta models at 0.75 days. Since the \elem[56]{Ni} mass is fixed, there are substantial variations in the mass of the IMEs. 
In particular, the mass of individual IMEs in model SUP is a factor of 3 to 5 larger than in model SUB2; model SUB1 has $\sim$1.3 to 1.4 times the mass of IMEs and $\sim$1/3 times the iron mass of SUB2. The initial iron abundance in SUB1 is almost a factor of 6 lower than in SUP. The mass of stable nickel (\elem[58]{Ni} and \elem[60]{Ni}) is 0.011, 0.026, 0.025, and 0.030 \Msun\ for SUB1, SUB2, CHAN, and SUP. The presence of stable nickel features in nebular spectra, and the ability to measure the nickel abundance, is discussed in Sections \ref{section_spec_evo} and \ref{section_problems}.

In \fig~\ref{mass_vs_v} we show the outward cumulative mass as a function of velocity. Less than 1 per cent of the mass lies beyond 25\,000 \kms\ for all models, and hence we restrict future model ejecta comparisons to velocities less than 30\,000\,\kms. Higher velocity material makes very minor contributions to synthetic spectra.

\begin{figure*} 
\begin{minipage}[t]{\linewidth}
\centering
\includegraphics[scale=0.8]{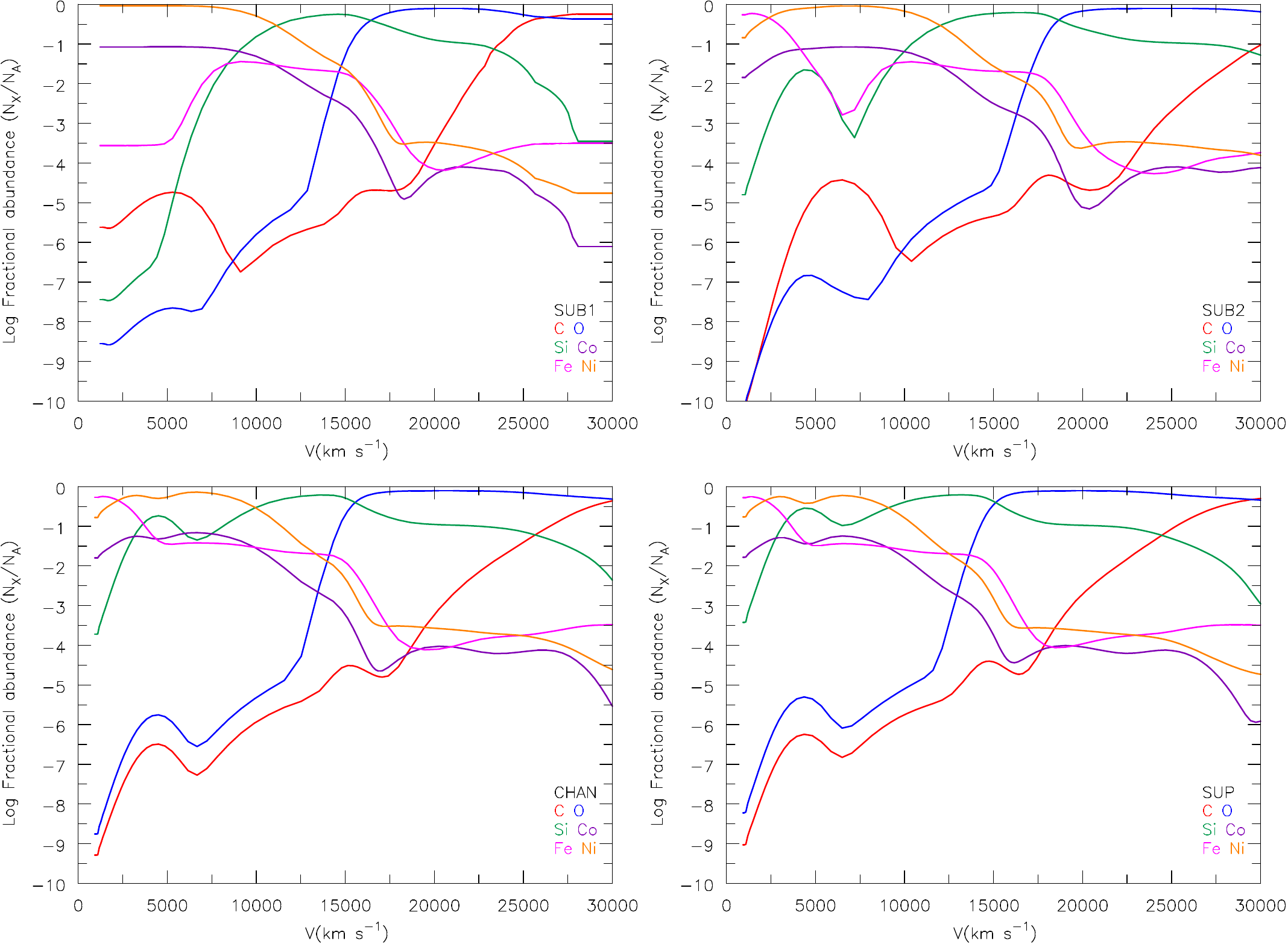}
\end{minipage}
\caption{The initial abundance as a function of velocity for carbon, oxygen, silicon, iron, cobalt, and nickel for model SUB1 (top left), SUB2 (top right), CHAN (bottom left), and SUP (bottom right). Notice the absence of the `\elem[56]{Ni} hole' in model SUB1 (i.e. presence of \elem[56]{Ni}), a result of having lower densities during the initial explosion. Also apparent is the  overlap of the nickel distribution with IMEs around 8000 to 12\,000 \kms\ in all models.}
\label{initial_rho_all}
\end{figure*}

\fig~\ref{initial_rho_all} compares the initial mass fraction at 0.75 days for all models. All but model SUB1 exhibit an `\elem[56]{Ni} hole' at velocities less than $\sim$2500\,\kms. As noted earlier, the hole in SUB2 is artificial, and arises since the model was scaled from a model based on the explosion of a \Mch\ WD which has a higher central density.% during nucleosynthesis.
  
\section{Light Curves}%%%%%%%%%%%%%%%%%%%%%%%%%%%%%%%%%%%%%%%%%%%%%%%%
\label{section_lc}
The light curves and colour evolution of Type Ia models depends on the progenitor system, and potentially offer a means to distinguish between progenitor systems. Due to differences in ejecta mass, the diffusion time varies between models, giving rise to morphological separations in both the width of each bolometric light curve and the peak luminosity. However, from work by \cite{Pinto_Eastman2000a,Pinto_Eastman2000b}, we expect this effect to be small.

\begin{figure*} 
\centering
\includegraphics[scale=1]{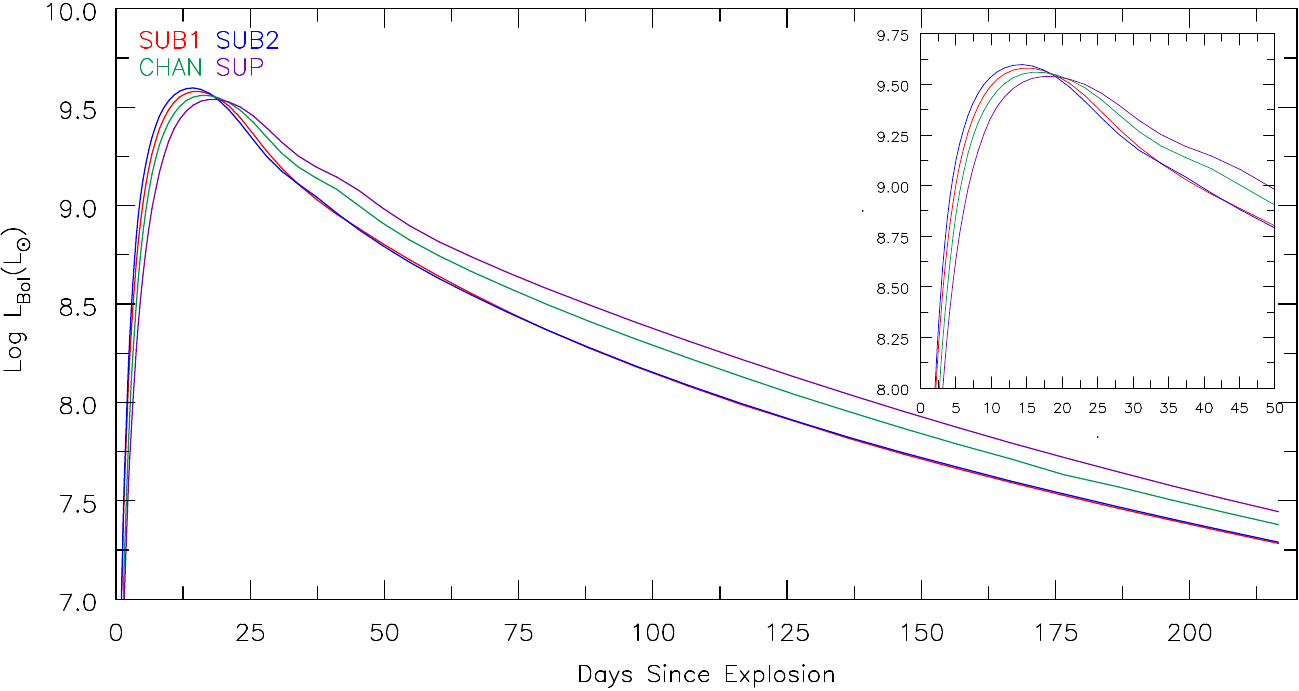}
\caption{The bolometric luminosities for all models as a function of age since explosion. Lower mass models evolve faster in their bolometric light curves compared with higher mass models -- the lower the ejecta mass, the earlier the time of bolometric maximum (Table \ref{lc_data}).}
\label{LogLbol}
\end{figure*}

\fig~\ref{LogLbol} shows the bolometric light curves of all models relative to the time of explosion. Decreasing ejecta mass (shorter diffusion time) corresponds to a faster evolving supernova. In days since explosion, the bolometric luminosity maximum occurs at 14.4 (3.80$\times10^9$ \Lsun), 14.4 (3.96$\times10^9$ \Lsun), 15.84 (3.63$\times10^9$ \Lsun), and 17.42 (3.47$\times10^9$ \Lsun) for models SUB1, SUB2, CHAN, and SUP respectively. To characterize the bolometric light curves, we list $t_{\rm max}$ (time of bolometric maximum), $t_{-1/2}$ and $t_{+1/2}$ (the times to rise from half bolometric maximum luminosity to maximum and to decline from bolometric maximum back to half of maximum -- see \cite{Contardo2000}) in Table \ref{lc_data}. Between $\sim$1--1.7 \Msun, we have roughly a 1 day difference in half light rise times ($t_{-1/2}$), with SUB ejecta models rising faster. After bolometric maximum, the SUP ejecta model takes roughly $\sim3.5$ days longer than the SUB ejecta models to decline to half light ($t_{+1/2}$). With precise measurement determinations of the rise time in $L_{\rm bol}$, we can put stronger constraints on the ejecta mass for a given \elem[56]{Ni} mass. 

As the \elem[56]{Ni} mass is the same for all models, differences in the bolometric light curve evolution are primarily due to ejecta mass. However, despite differences in ejecta mass of $\sim$70 per cent, the differences are relatively small, though measurable. This effect of ejecta mass should be clearly visible from a statistical sample of deep high cadence observations of SNe Ia. 

\begin{figure*} 
\hspace*{-0.75cm}\includegraphics[scale=0.8]{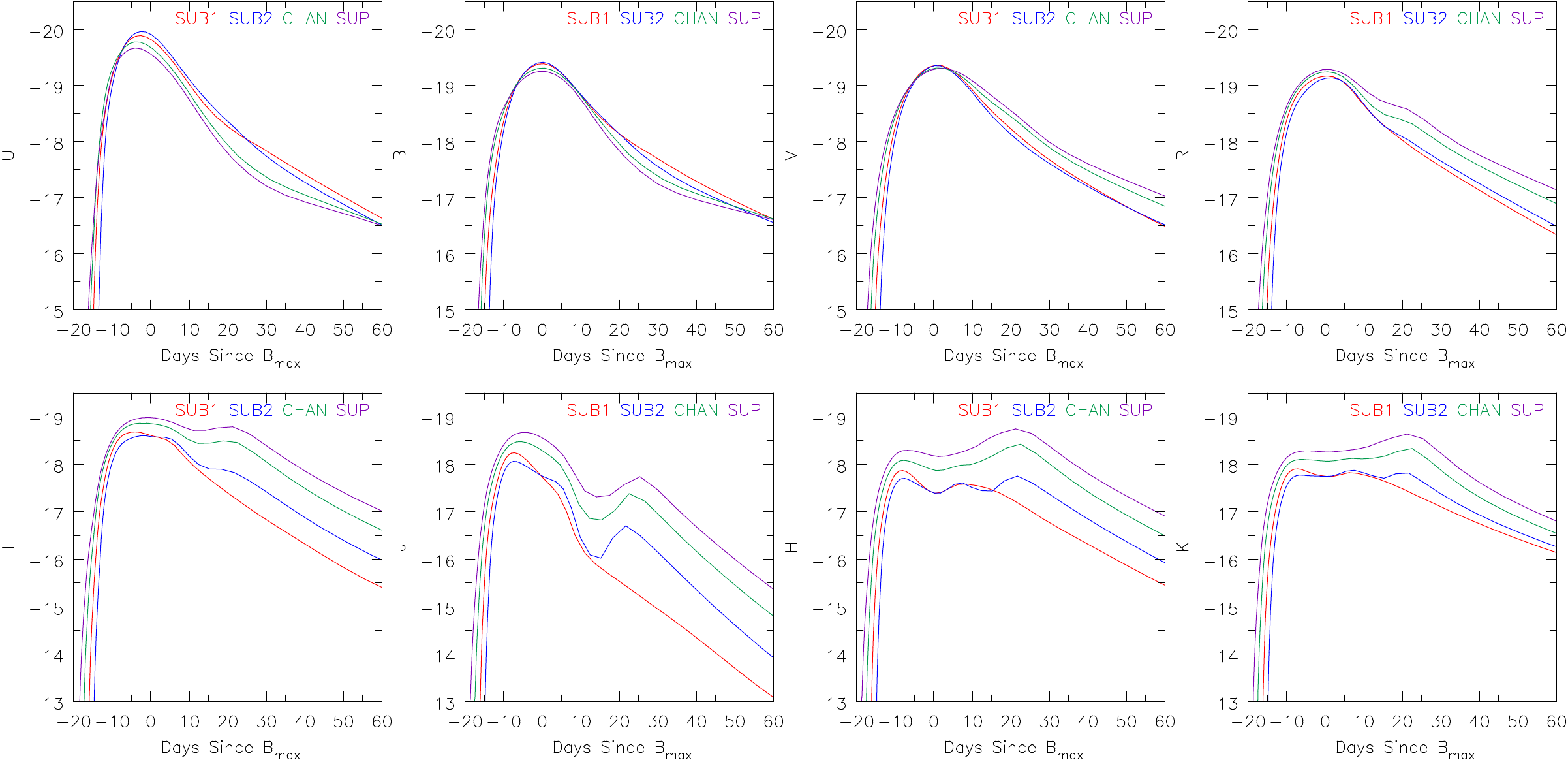}
\caption[]{The plots show all LCs as a function of age since $B$-band maximum. Unlike the bolometric luminosity (\fig~\ref{LogLbol}), the $B$-band LCs for SUB1 and SUB2 are different -- SUB1 is roughly 0.2 magnitudes brighter than SUB2 between $\sim$25-60 days post $B$-band maximum. This difference in evolution arises because SUB1 lacks the `\elem[56]{Ni} hole' (\fig~\ref{initial_rho_all}), and hence is hotter, and has higher ionization, at lower velocities than SUB2. The $H$-band shows a triple peak for model SUB2 as the ionization shifts downwards, enhancing charge 2+ and then 1+ ion abundances of IGEs like cobalt and iron.}
\label{lc_wrtb}
\end{figure*}

\fig~\ref{lc_wrtb} shows synthetic light curves plotted in days since $B$-band maximum. The light curves for the two sub-\Mch\ ejecta models tend to be more luminous at maximum (with the exception of the NIR bands) but have fainter nebular luminosities. The former arises because at bolometric maximum $L_{\rm bol}\approx L_{\rm decay}$ and since the peak is earlier, the decay rate is greater. The latter arises because of the less efficient trapping of \grays. Table \ref{lc_data} also lists the peak LC absolute magnitudes for different Johnson bands ($M_U\rightarrow M_K$), rise times (in days since explosion), and decline parameter $\Delta M_{15}(X)$ for a given band $X$. % \citep[which is the change in $X$-band magnitude 15 days after $X$-band maximum;][]{Phillips1993}.
All bands show a faster rise time for lower-mass ejecta models, just as they do for $L_{\rm bol}$.  

Observational evidence suggests that $H$-band photometry for SNe Ia can provide higher accuracy than the $B$-band calibration galactic distances \citep{Krisciunas2004,Krisciunas2007,Wood-Vasey2008}. However, our models have a spread of almost one magnitude in the $H$-band at the time of $B$-band maximum light. Following the temperature separation between models (higher mass $\rightarrow$ cooler ejecta), the higher mass ejecta also show larger flux in the $H$-band at all epochs. Our light curves still show two peaks in the $H$-band as normal SNe Ia do, a consequence of the ionization shift in iron (and other IGEs) going from ionization states 2+$\rightarrow$1+ \citep{Kasen2006}. The second peak is of the same brightness for our SUB models compared to higher ejecta mass models. Past the second $H$-band peak, we see roughly a constant decline in all models until close to 150 days past maximum light. 


\begin{figure*} 
\includegraphics[scale=0.9]{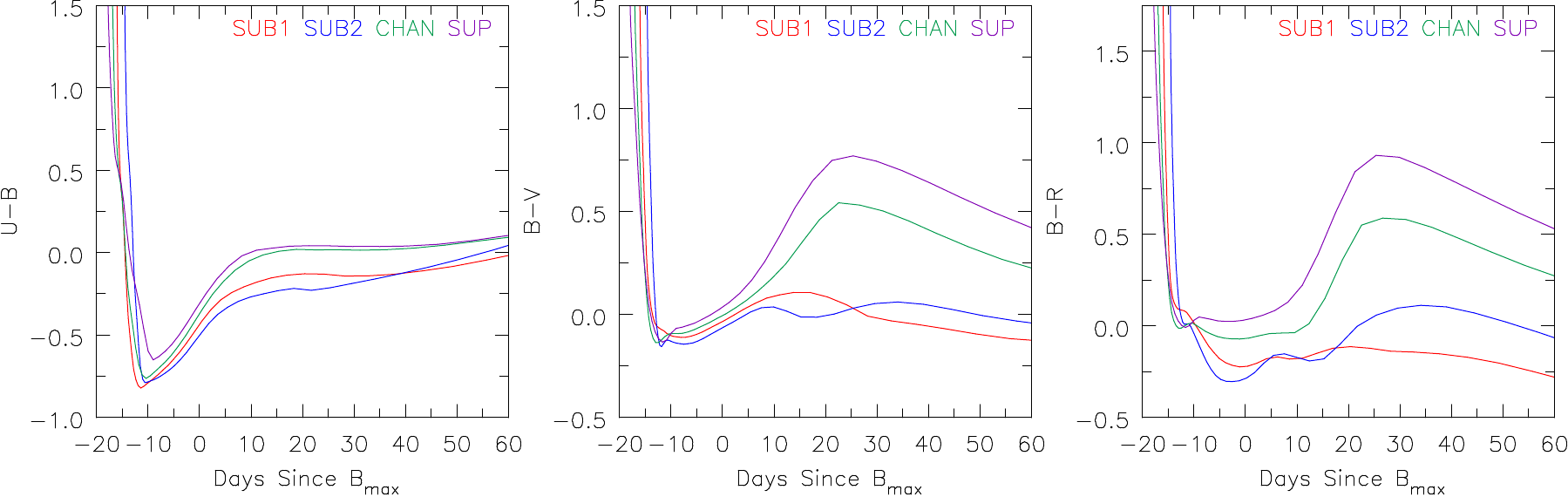}
\caption{The $U-B$, $B-V$, and $B-R$ colours relative to $B$-band maximum. Models SUB1 and SUB2 show bluer colours for all epochs. Ten days before $B$-band maximum all models have similar colours. As the ejecta ages, differences in the $B-R$ colour become much larger -- at $B$-band maximum, the $B-R$ colour differs by  $\sim$0.3 mag between SUB2 and SUP, while at later epochs the colour difference between SUB1 and SUP can be in excess of 1 magnitude.}
\label{LCcolor}
\end{figure*}

\begin{figure*} 
\begin{minipage}[t]{\linewidth}
\centering
\includegraphics[scale=0.9]{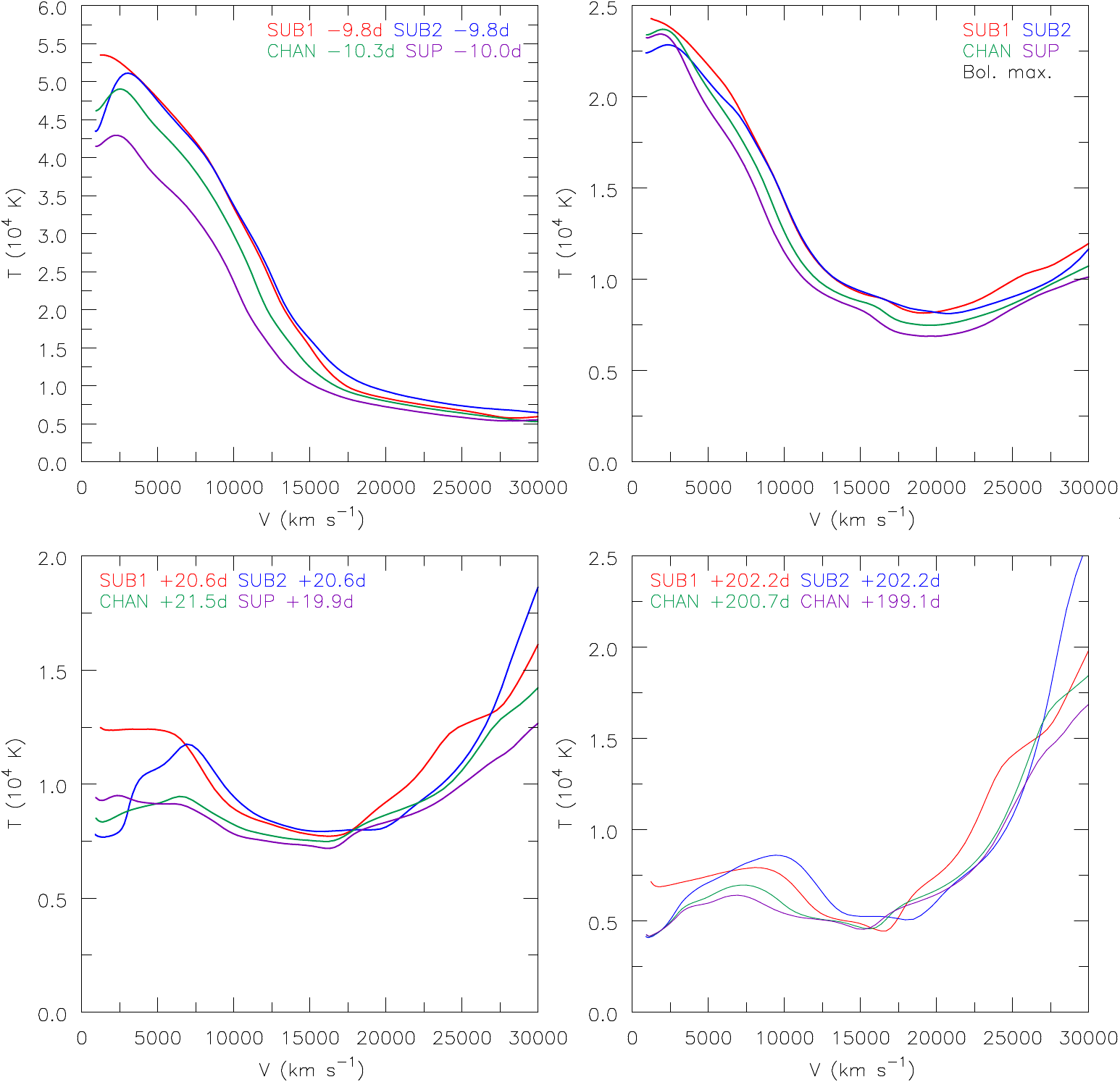}
\end{minipage}
\caption{Temperature evolution as a function of velocity for models SUB1, SUB2, CHAN, and SUP at epochs $-10$, $+0$, $+20$, and $+200$ days relative to bolometric maximum. SUB1 exhibits higher temperatures at low velocities compared to SUB2, CHAN and SUP -- the higher temperature arises because of the much higher \elem[56]{Ni} abundance in the innermost ejecta of SUB1 (\fig~\ref{initial_rho_all}).
}
\label{TvsV_all}
\end{figure*}

\begin{figure*}
\begin{minipage}[t]{\linewidth}
\centering
\hspace*{-0.9cm}
\includegraphics[scale=1.1]{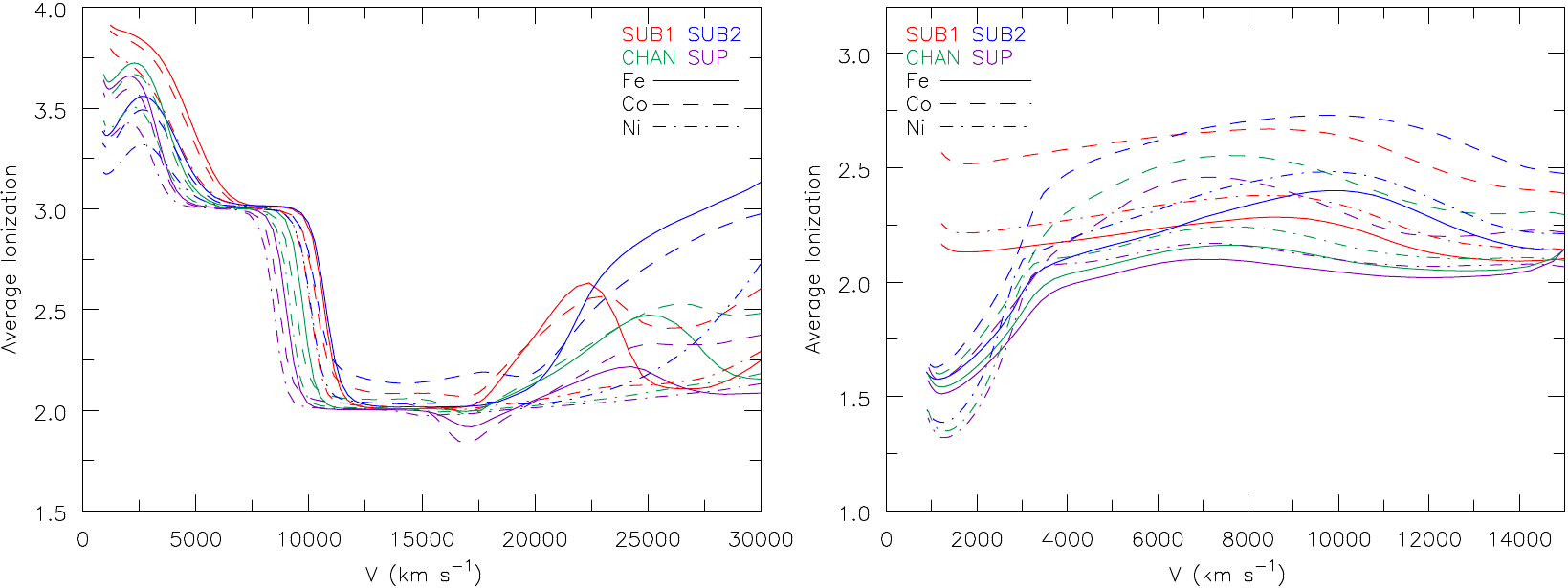}
\end{minipage}
\caption{Average ionization state for bolometric maximum (left) and 216 days post explosion (right) for IGEs. The average ionization is defined as the total number of free electrons from each species divided by the total species population. At bolometric maximum, the average ionization rises beyond 20\,000~\kms\ due to the low densities inhibiting recombination. The average ionization state is higher for sub-\Mch\ models, a result of more heating per unit mass. At nebular times, SUB1 remains at a higher average ionization in the inner region due to a larger deposition of energy by positrons arising from the higher \elem[56]{Co} mass fraction in the inner region (see \fig~\ref{edep_ratio}).}
\label{IF_frac}
\end{figure*}

\fig~\ref{LCcolor} shows the $U-B$, $B-V$, and $B-R$ colour evolution relative to $B$-band maximum. It shows that the lower mass models are bluer at most epochs. There is over a magnitude difference in $B-R$ from sub-\Mch\ to super-\Mch\ around 25 days after $B_{\rm max}$. SUB models are bluest in color post $B$-band maximum compared to higher mass models. Post maximum, model SUP remains the most red of all the models, while SUB1 remains bluer than other models after $+40$ days, a result of higher temperatures and ionization due to larger $M(^{56}{\rm Ni})/M_{\rm ej}$. This is explored in \figs~\ref{TvsV_all} and \ref{IF_frac} and then in Section \ref{section_spec_evo} where we discuss the spectral evolution. 

\begin{figure} 
\begin{minipage}[t]{\linewidth}
\centering
\includegraphics[scale=0.9]{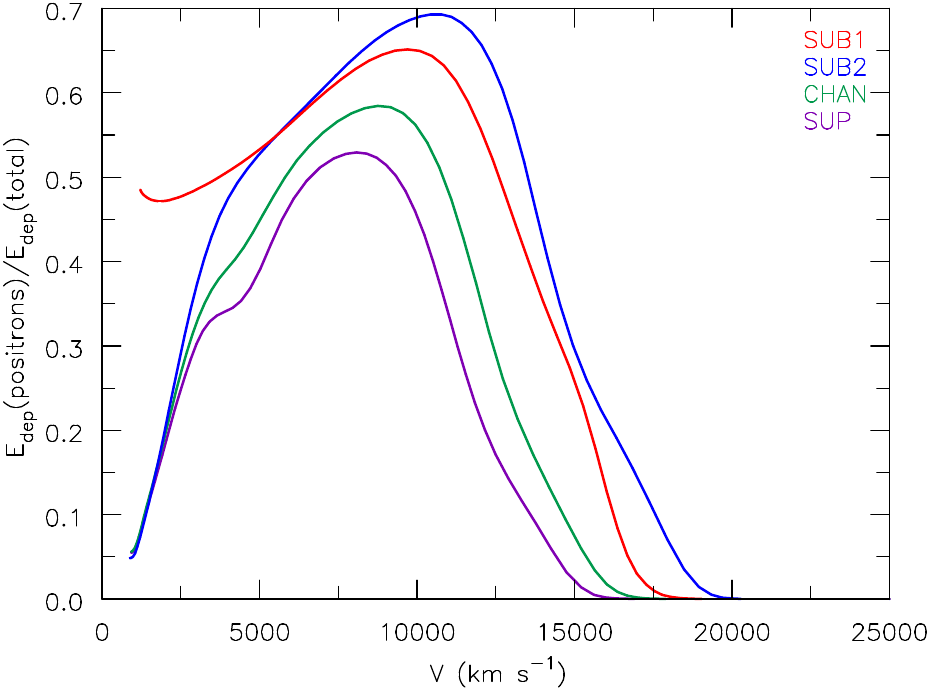}
\end{minipage}
\caption{Ratio of the locally deposited energy from positrons (mainly from \elem[56]{Co}) to the energy deposited by nuclear decays in the ejecta at about 216 days post explosion. SUB1 shows a much higher ratio shortward of 5000~\kms\ due to the lack of a `\elem[56]{Ni} hole' and higher \elem[56]{Ni} production in the core. This leads to higher ionization and a higher temperature in the inner region (in combination with lower densities) compared to other models.}
\label{edep_ratio}
\end{figure}

\fig~\ref{TvsV_all} shows the temperature evolution of our models, with SUB1 and SUB2 maintaining higher temperatures at all epochs. \fig~\ref{IF_frac} shows the average ionization for a few IGEs such as iron, cobalt, and nickel. SUB models show a higher ionization compared to higher mass models, producing ejecta with bluer colors. The monotonic temperature distribution of SUB1 below 5000 \kms\ is due to the lack of a `\elem[56]{Ni} hole'. \fig~\ref{edep_ratio} shows the fraction of the energy deposition from positrons to the total energy deposition at about 216 days post explosion. At late times SUB1 maintains a higher ionization in the innermost ejecta due to more assumed local positron energy deposition. Generally speaking, SUB1 and SUB2 have higher temperatures at comparable epochs than those of CHAN and SUP. We further discuss the implications of higher temperatures in Section \ref{section_spec_evo}.

\begin{table*}
\begin{minipage}[t]{\linewidth}
\hspace*{-0.75cm}
 \begin{tabular}{| l  | c | c | c | c | c | c | c | c | c | c | c | c |} 
 \hline
 Model & $M_U$ & $t(U_{\rm max})$ & $\Delta M_{15}(U)$ & $M_B$ & $t(B_{\rm max})$ & $\Delta M_{15}(B)$ & $M_V$ & $t(V_{\rm max})$ & $\Delta M_{15}(V)$ & $M_R$ & $t(R_{\rm max})$ & $\Delta M_{15}(R)$ \\
 & (mag) & (days) & (mag) & (mag) & (days) & (mag)  & (mag) & (days) & (mag) & (mag) & (days) & (mag)  \\ [0.5ex]  \hline\hline
SUB1 & $-$19.89  & 14.13  &  1.07  & $-$19.39  & 16.97  &  0.95  & $-$19.36  & 17.99  &  0.88  & $-$19.17  & 17.40  &  0.89  \\
SUB2 & $-$19.96  & 13.35  &  1.08  & $-$19.41  & 15.64  &  0.94  & $-$19.36  & 16.48  &  0.96  & $-$19.14  & 17.01  &  0.91  \\
CHAN & $-$19.78  & 14.90  &  1.06  & $-$19.31  & 18.54  &  0.94  & $-$19.31  & 19.91  &  0.68  & $-$19.24  & 18.92  &  0.76  \\
SUP & $-$19.67  & 15.92  &  1.03  & $-$19.25  & 19.86  &  0.99  & $-$19.31  & 21.89  &  0.60  & $-$19.29  & 20.48  &  0.59  \\ \hline
& $M_I$ & $t(I_{\rm max})$ & $\Delta M_{15}(I)$ & $M_J$ & $t(J_{\rm max})$ & $\Delta M_{15}(J)$ & $M_H$ & $t(H_{\rm max})$ & $\Delta M_{15}(H)$ & $M_K$ & $t(K_{\rm max})$ & $\Delta M_{15}(K)$ \\
& (mag) & (days) & (mag) & (mag) & (days) & (mag) & (mag) & (days) & (mag) & (mag) & (days) & (mag) \\ \hline\hline
SUB1 & $-$18.68  & 12.86  &  0.70  & $-$18.24  &  9.60  &  1.54  & $-$17.87  &  8.76  &  0.29  & $-$17.90  &  9.62  &  0.08  \\
SUB2 & $-$18.60  & 13.76  &  0.64  & $-$18.06  &  8.39  &  1.03  & $-$17.70  &  7.81  &  0.09  & $-$17.77  &  9.27  & $-$0.08  \\
CHAN & $-$18.87  & 16.41  &  0.44  & $-$18.48  & 12.73  &  1.26  & $-$18.08  & 10.73  &  0.10  & $-$18.10  & 12.40  & $-$0.02  \\
SUP & $-$18.99  & 19.01  &  0.27  & $-$18.67  & 15.29  &  1.17  & $-$18.30  & 12.91  &  0.02  & $-$18.28  & 15.25  & $-$0.07  \\ \hline
& $t_{-1/2}$ & $t_{+1/2}$ & $t_{\rm max}$ & $L_{\rm bol}$ \\
& (mag) & (days) & (mag) & $(10^9$\Lsun) \\ \hline\hline %\cline{1-5}
SUB1 & 8.18 & 13.0 & 14.4 & 3.80 \\
SUB2 & 8.02 & 12.44 & 14.4 & 3.96\\
CHAN & 8.83 & 14.75 & 15.84 & 3.63 \\
SUP & 9.27 & 16.57 & 17.42 & 3.47 \\ \hline%\cline{1-5}
 \end{tabular}
\caption[]{Light curve parameters such as absolute magnitude maximum ($M_X$), time of maximum ($t(X_{\rm max})$), and decline parameter ($\Delta M_{15}(X)$) for band $X$. Also included are half light rise and decline times for $L_{\rm bol}$ ($t_{-1/2}$ and $t_{+1/2}$), time of bolometric maximum ($t_{\rm max}$), and maximum bolometric light ($L_{\rm bol}$).}
\label{lc_data}
\end{minipage}
\end{table*}

\section{Spectral Diagnostics}\label{section_spec_evo} %%%%%%%%%%%%%%%%%%%%%%%%%%%%%%%%%%%%%%%
\begin{figure*} 
\begin{minipage}[t]{\linewidth}
\centering
\includegraphics[scale=1]{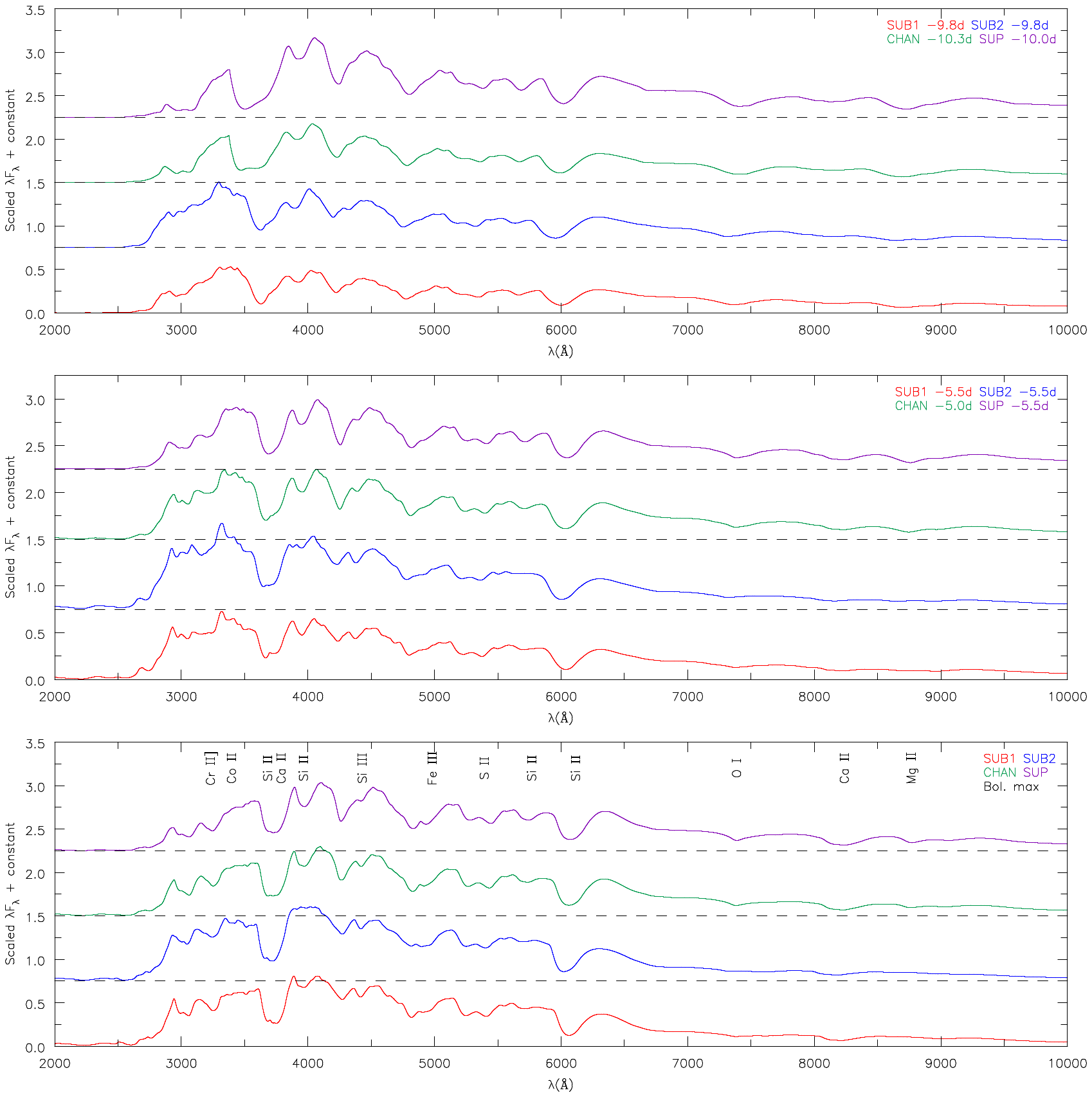}
\end{minipage}
\caption{Spectral comparisons between models in the pre-maximum ($-10$ and $-5$ days relative to bolometric maximum) and bolometric maximum plotted in $\lambda F_\lambda$ vs \lb\ in order to contrast the NIR tail of the spectrum. All models have been scaled by the same factor, and we have added an offset to allow spectra to be more easily distinguished. During these phases the optical spectra (3500 to 7000 \AA) are very similar. However, there are differences in the amount of UV blanketing occurring shortward of 3500 \AA, and in the strength of \oi\ and the \caii\ NIR triplet. We also see that the strength of the \oi\ feature decreases with age while the \caii\ NIR triplet increases in strength.}
\label{premax_max_spec}
\end{figure*}

Spectra provide important constraints for distinguishing progenitor and explosion models. However, despite a 70 per cent difference in mass, model optical spectra (\figs~\ref{premax_max_spec}, \ref{postmax_1}, \ref{postmax_2}, \ref{nebular_1}, and \ref{nebular_2}) at most phases are similar, consistent with work by \cite{Blondin2013} who concluded that SNe Ia are mainly distinguished by their \elem[56]{Ni} mass. This similarity is  observed for classical SNe Ia where differences in optical spectra are generally rather subtle \citep{Filippenko1997}, which presumably occurs because of similarities in composition. Interestingly, optical spectra of models SUB1 and SUB2 are remarkably similar for most phases of evolution, despite the different methods to produce these ejecta models.  However, there are differences between models, particularly in the infrared, and these do give rise to useful diagnostics. We investigate these diagnostics by comparing model spectra at  time steps of approximately $-10$, $-5$, $0$, $+5$, $+10$, $+20$, $+50$, $+100$, and $+200$ days from bolometric maximum (\figs~\ref{premax_max_spec}, \ref{postmax_1}, \ref{postmax_2}, \ref{nebular_1}, and \ref{nebular_2}). Thus, when comparing observational spectra of SNe Ia from comparable \elem[56]{Ni} mass, these diagnostics will separate events by ejecta mass. Note that model spectra are plotted for vacuum wavelengths; however, wavelengths $\geq$2000 \AA\ listed are quoted in air.

\fig~\ref{premax_max_spec} shows the early spectral evolution ($-10$, $-5$, and 0 days relative to bolometric maximum) plotted in $\lambda F_{\lambda}$ (arbitrary units). We label the contributions of important species at bolometric maximum. \figs~\ref{postmax_1} and \ref{postmax_2} show the post maximum photospheric phase ($+5$, $+10$, $+20$, $+50$) plotted in $\lambda F_{\lambda}$ (arbitrary units). \figs~\ref{nebular_1} and \ref{nebular_2} show nebular spectra at $+100$ and $+200$ days post bolometric maximum plotted in $F_{\lambda}$ (arbitrary units). Contributions from important species are labelled. Notice the transition towards predominantly forbidden lines in nebular spectra. These figures are used to highlight important diagnostics.

\subsection{[\nkii] 1.939 $\mu$m}\label{nickel_line_diag}%%%%%%%%%%%
In SNe Ia, the nickel abundance is sensitive to the progenitor mass and/or explosion scenario. In 1D explosion modeling, higher central densities have higher neutronisation that leads to more stable \elem[58]{Ni} being produced during nuclear burning \citep{Nomoto1984,Khokhlov1991a,Khokhlov1991b}. This 1D modeling implies that sub-\Mch\ SNe Ia will show a lower abundance of \elem[58]{Ni} compared to \Mch\ SNe Ia (for the same \elem[56]{Ni} mass). However, 3D DDT modeling suggests that the \elem[56]{Ni} hole predicted in 1D \Mch\ WD DDT models may be absent, and both \elem[56]{Ni} and \elem[58]{Ni} extend from the lowest velocities to about 10\,000 \kms \citep{Kasen2009,Seitenzahl2013}. \elem[22]{Ne} settling in sub-\Mch\ has also been proposed as a way to enhance the neutronisation; however, the time-scale for gravitational settling can be $\sim10^9-10^{10}$ yrs \citep{Bildsten2001}. Therefore, nickel diagnostics, particularly at nebular times, may constrain the progenitor scenario, nucleosynthesis, and explosion mechanism \citep{Woosley1997,Iwamoto1999,Stehle2005,Mazzali2006,Gerardy2007,Maeda2010,Mazzali2011,Mazzali2012,Mazzali2015}. At nebular times most of the \elem[56]{Ni} will have decayed, and any nickel emission features are due to stable nickel, and in particular \elem[58]{Ni} and \elem[60]{Ni}, which are expected to be underabundant in (1D) sub-\Mch\ DDT models compared with \Mch\ models. The width of any observed nebular nickel feature will constrain the hydrodynamic width of the emitting region, thus testing model predictions about the presence of a \elem[56]{Ni} hole. Therefore, nickel features may offer the best diagnostic for ejecta masses below \Mch\ if ejecta do or do not have \elem[56]{Ni} holes as predicted by 1D modeling. 

In the optical nebular spectra at $+100$ and $+200$ days (\figs~\ref{nebular_1} and \ref{nebular_2}), the [\nkii] \lb\lb7378,7412 lines are blended, and hence not very useful for abundance determinations. However, in the NIR there is a  forbidden [\nkii] transition (3d8($^3$F)4s${}^2$\,F$_{7/2}\,-\,$3d8($^3$F)4s$^4$\,F$_{9/2}$) at 1.939 $\mu$m which in our synthetic spectra is relatively blend-free. It overlaps with telluric lines in low-redshift SNe, but higher redshift ($z>0.08$) SNe avoid telluric absorption. While observations of this line appear to be rare, \cite{Friesen2014} find evidence for this line in spectra of SN2011fe, SN2014J, and SN2003du.

In our models, SUB1 shows no evidence of [\nkii] 1.939 $\mu$m. On the other hand, SUB2, CHAN, and SUP show the line, with a strength that correlates with ejecta mass. The absence of [\nkii] 1.939 $\mu$m in SUB1 arises from two effects --- SUB1 has a smaller amount of stable nickel (see Table \ref{model_info_abund}) and a higher ionization than the other models. The presence of \elem[56]{Co} under $\sim$3000\,\kms\ in SUB1 means that there is a great amount of heating from positrons, which deposit their energy locally (\fig~\ref{edep_ratio}). This, combined with the lower densities, leads to both a higher temperature, and a higher ionization (\figs~\ref{TvsV_all} and \ref{IF_frac}). Surprisingly, and despite their similar ionization potentials, \feii\ cannot be used as an ionization tracer for \nkii. In SUB1 \nkiii/\nkii\ is significantly larger than \feiii/\feii. This arises because the photoionization of \nkii\ is dominated by large resonances in its photoionization cross-section. Therefore, the absence of [\nkii] 1.939 $\mu$m in SN Ia spectra at 100--200 days indicates that the mass of the progenitor is below \Mch. However, we re-emphasize that SUB2 is a scaled \Mch\ model, so it is not a true sub-\Mch\ model. SUB2 has a stable nickel core and shows [\nkii] 1.939 $\mu$m.

\subsection{Ionization}%%%%%%%%%%%%%%%%%%%%%%
Once the \elem[56]{Ni} mass is determined via ``Arnett's rule", which states that the luminosity at bolometric maximum is equal to the \elem[56]{Ni} decay chain luminosity, or using LC fitting like that of \cite{Scalzo2014a,Scalzo2014b}, one can separate different SNe Ia based on ejecta mass using differences in ionization/temperature (see \fig~\ref{TvsV_all}). This result follows from the heating per gram available to the gas. The greater the ejecta mass is, the lower the heating rate per gram is. Consequently, ejecta with a larger $M(^{56}{\rm Ni})/M_{\rm ej}$ are hotter -- see \cite{Blondin2017} for $\dot{e}_{\rm decay}\equiv L_{\rm decay}/M_{\rm tot}$. Indeed our models indicate that low mass WD models, for a given \elem[56]{Ni} mass, maintain higher ionizations throughout their spectral evolution. 

\subsubsection{UV-Blanketing}%%%%%%%%%%%%%%%%%%%%
Previous studies focused on the UV variability and used the UV spectral region for understanding SNe Ia. These studies looked at the role of metallicity on UV blanketing \citep{Lentz2000,Walker2012,Wang2012,Foley2013} finding that lower metallicities shift the blanketing blueward. \cite{Foley2016} looked at a sample of SNe Ia and found the UV diversity linearly correlates with the optical LC shape. In particular, the strength of UV line flux measurements ($\sim$2030 \& 2535 \AA) increases with increasing $\Delta m_{15}(B)$. Other studies of UV variation hope to use it as a cosmological utility \citep{Ellis2008,Sullivan2009} to improve standardizability. 
Therefore, understanding how $M_{\rm ej}$ for a given \elem[56]{Ni} mass influences the UV spectrum is important to the astronomical community.

Until the ejecta begins entering its nebular phase ($\sim$100 days), we see larger UV blanketing shortward of 4000 \AA\ for larger mass ejecta. This effect is attributed to a temperature difference between models. We see in \figs~\ref{TvsV_all} and \ref{IF_frac} that below 25\,000 \kms models with higher temperatures have higher ionizations, seen as a shift in the line blanketing to higher frequencies. Pre-maximum spectra show the \caii\ feature (H \& K lines near \lb3500) is affected by UV blanketing, making it difficult to distinguish in SUP and CHAN (\fig~\ref{premax_max_spec}). \tkii\ contributes to much of the blanketing more than 5 days before maximum, while \tkii, \feii, and \feiii\ shape the UV spectra just prior to maximum. Around maximum, \coii\ contributes much of the UV blanketing (below 3500 \AA) with the strongest blanketing occurring in model SUP. Looking inwards of 25\,000 \kms, SUB1 and SUB2 show a higher ionization of cobalt than that of models SUP and CHAN. For \coii\ there is about a half dex difference in ionization between SUB1 and SUP. These \coii\ differences show up as absorption affecting the slope of the feature at $\sim3500$ \AA. Post-maximum (\figs~\ref{postmax_1} and \ref{postmax_2}), there is less variation in UV blanketing between the models.

If we compare the peak fluxes at bolometric maximum (\fig~\ref{premax_max_spec}) of three UV features (namely the features near $\sim$2850\,\AA, $\sim$3150\,\AA, and $\sim$3550\,\AA), we can characterize the level of blanketing by comparing the flux at peak in each feature. For all ejecta models, the flux ratio $F(3150)/F(2850)$ is close to unity (0.93, 1.07, 1.03, and 1.11 for SUB1, SUB2, CHAN, and SUP). However, comparing these lines to the feature just short of the \caii\ H\&K and \skii\ $\sim$\lb3660 \AA\ absorption profile, we see that the flux ratio $F(3550)/F(2850)$ is strongly dependent on ejecta mass. This flux ratio $F(3550)/F(2850)$ is 0.99, 1.01, 1.20, and 1.75 for SUB1, SUB2, CHAN, and SUP. These UV features reflect the temperature and ionization of the ejecta and offer a diagnostic of ejecta mass for a given \elem[56]{Ni} mass. 
\subsubsection{Optical and IR} \label{optical_NIR}
Besides variations in UV blanketing, other ionization diagnostics are seen in optical and infrared spectra. For instance, leading up to maximum (\fig~\ref{premax_max_spec}), each model shows a different strength of the \skiii\ triplet (\lb\lb4553,4568,4575) absorption, which is strongest in the models SUB1 and SUB2. SNe Ia typically classified as normal, such as SN2011fe, show the \skiii\ feature around 4400 \AA\  \citep{Pereira2013} as our model CHAN does. SNe Ia such as SN2003hv, thought to be a sub-\Mch\ event \citep{Mazzali2011}, show this absorption feature much more strongly \citep{Leloudas2009}, as in our SUB models. 

Post maximum (\fig~\ref{postmax_2}), the near-infrared part of the spectrum begins to show prominent permitted \feii\ (9997.58, 10501.50, 10862.64, 16787.18, and 16873.20 \AA) and \coii\ features (11829.72, 15758.43, 16064.48, 16360.46, 16687.30, 21344.70, 21503.28, 22202.92, 22475.63, and 23612.53 \AA), as well as forbidden [\feiii] (22178.21, 22420.43, and 23478.80 \AA) and [\coiii] (12724.19, 15483.56, 17408.66, 19575.24, 20022.57, and 20973.15 \AA) lines. Many of the \coii\ and \feii\ features are absent in SUB1, a result of the higher ionization. 

Optical nebular spectra typically exhibit emission lines of \feiii\ and \coiii\ (\figs~\ref{nebular_1} and \ref{nebular_2}). In the NIR S$^{2+}$ and Ar$^{2+}$ show up in our model spectra as [\siii] \lb\lb9068,9530, and [\ariii] \lb\lb7135,7751, with the strength of these features relative to [\feiii] \lb4658 correlating with higher ejecta mass. As will be discussed in Section~\ref{section_problems}, our model spectra tend to exhibit too high an ionization, especially after 40 days. In particular, they lack strong [\feii] (e.g., [\feii] $\sim$4350 \AA). However [\feii] and [\coii] features are readily identified in the IR, except for model SUB1. Since SUB1 comes from the explosion of a sub-\Mch\ WD, its inner density is lower throughout its evolution compared to SUB2, and this hinders recombination. Further, SUB1 lacks the `\elem[56]{Ni} hole' seen in the later models, and hence the temperature in the inner region is higher than in the other models (see \figs~\ref{TvsV_all}, \ref{IF_frac}, and \ref{edep_ratio}).
\subsection{C/O and IMEs}
As the mass of C/O and IMEs is strongly correlated (by design from the density scalings) with the ejecta mass for a given \elem[56]{Ni} mass (see Table \ref{model_info_abund}), one should expect that lines from C/O and IMEs will provide a useful diagnostic tool for ejecta mass. As to be expected, our models show stronger absorption features for oxygen and IMEs for increasing ejecta mass. For example, the strength of absorption due to the \oi\ \lb\lb7772,7774,7775 triplet absorption correlates with ejecta mass in pre-maximum spectrum (\fig~\ref{premax_max_spec}). 
The feature fades by a few weeks post bolometric maximum. \mgii\ \lb\lb9218,9244 is another feature whose strength correlates with  high ejecta mass--see Table \ref{Sk2_widths} which lists the pseudo-equivalent widths (pEWs) measured by a straight line across the maxima of the absorption profile; it also  fades within a few weeks post bolometric maximum.

\begin{figure*} 
\begin{minipage}[t]{\linewidth}
\centering
\includegraphics[scale=1]{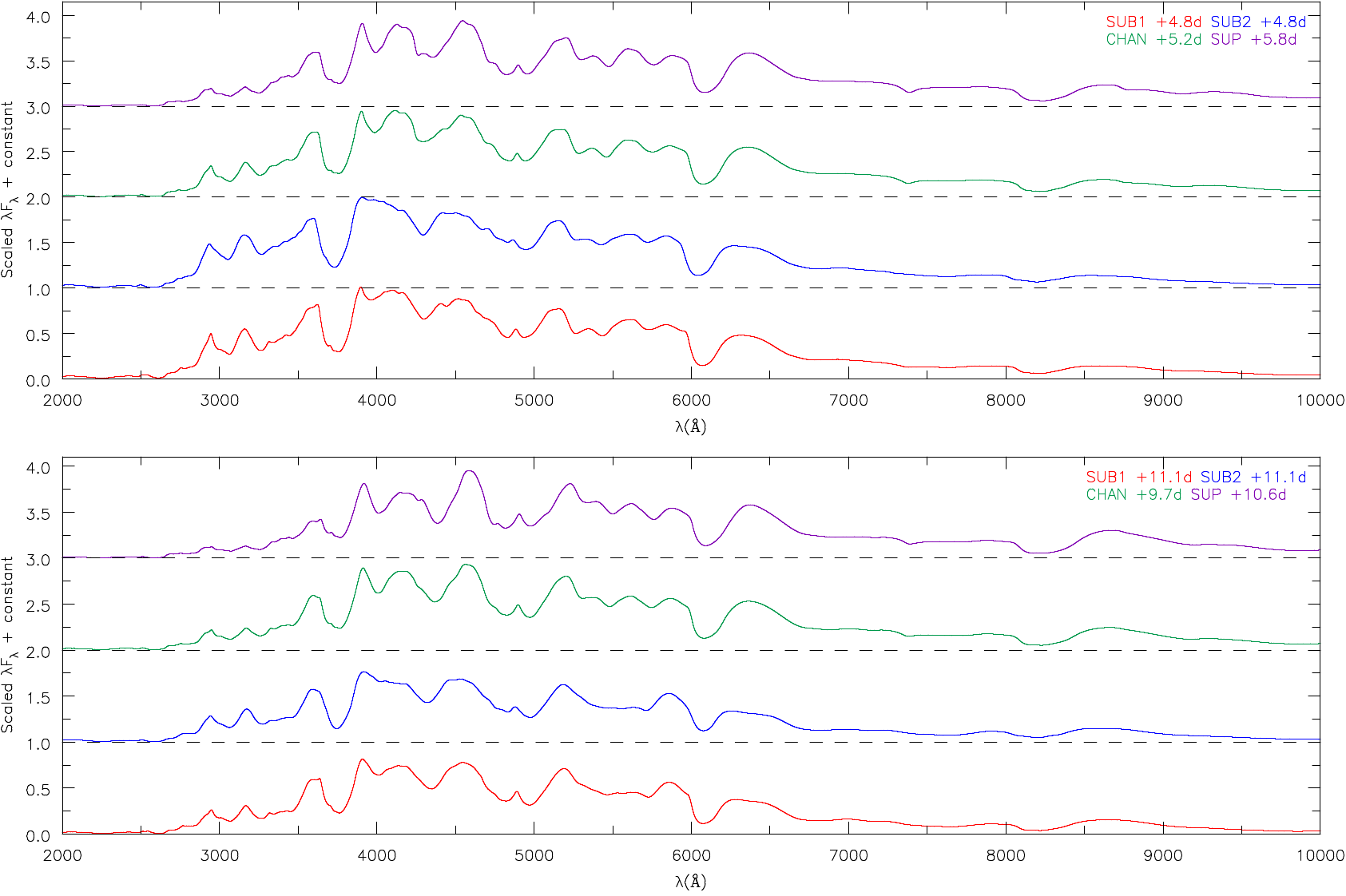}
\end{minipage}
\caption{Spectral comparisons between models in the post-maximum phases ($+5$ and $+10$ days) relative to bolometric maximum plotted in $\lambda F_\lambda$ vs \lb\ space in order to contrast the NIR tail of the spectrum. In each plot, all models have been scaled by the same value, and we have added a multiple of the offset to more easily distinguish the models. The optical spectra is again similar showing still the strong \skii\ doublet at $\sim$6100 \AA. The UV spectrum shows stronger blanketing in higher mass models. The strength of the \caii\ NIR triplet deviates between the models, showing stronger absorption/emission in the cooler (higher mass) models.}
\label{postmax_1}
\end{figure*}

\begin{figure*} 
\begin{minipage}[t]{\linewidth}
\centering
\includegraphics[scale=0.97]{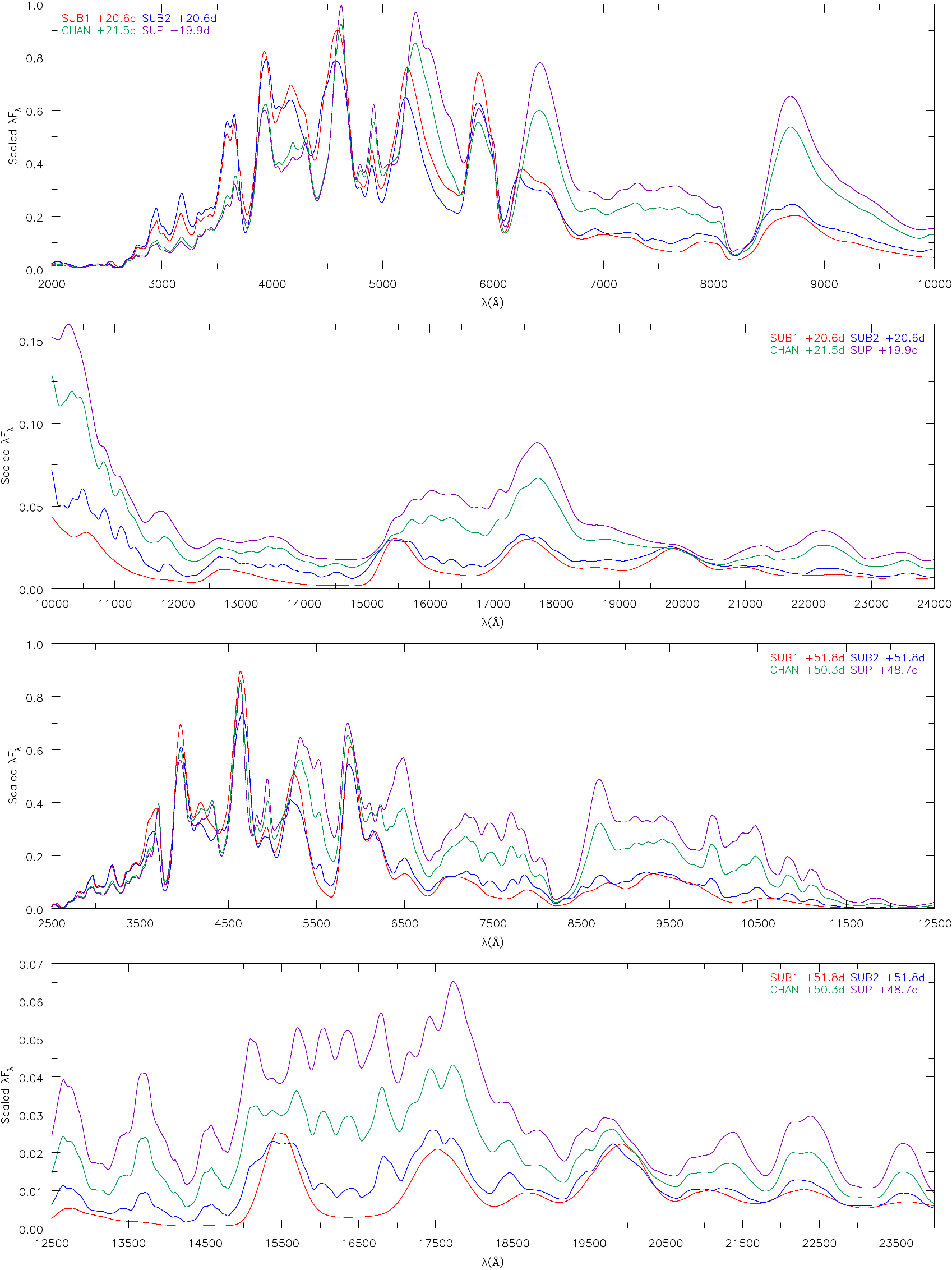}
\end{minipage}
\vspace{-0.5cm}
\caption{Spectral comparisons between models in the post-maximum phases at roughly $+20$ and $+50$ days relative to bolometric maximum. Each separate plot shows spectra scaled by the same value. At around $+20$ days, we continue to see a relatively homogeneous class of objects in the optical but stronger differences emerge in the NIR, such as the strength of the \caii\ NIR triplet and the growth of \feii\ and \coii\ lines -- see Section \ref{optical_NIR}. Despite the lack of \feii\ and \coii\ features in SUB1 (a consequence of having lower densities and higher temperatures/ionization in the inner ejecta), spectra of SUB1 and SUB2 are still fairly similar.}
\label{postmax_2}
\end{figure*}

\begin{figure*} 
\begin{minipage}[t]{\linewidth}
\centering
\includegraphics[scale=1.0]{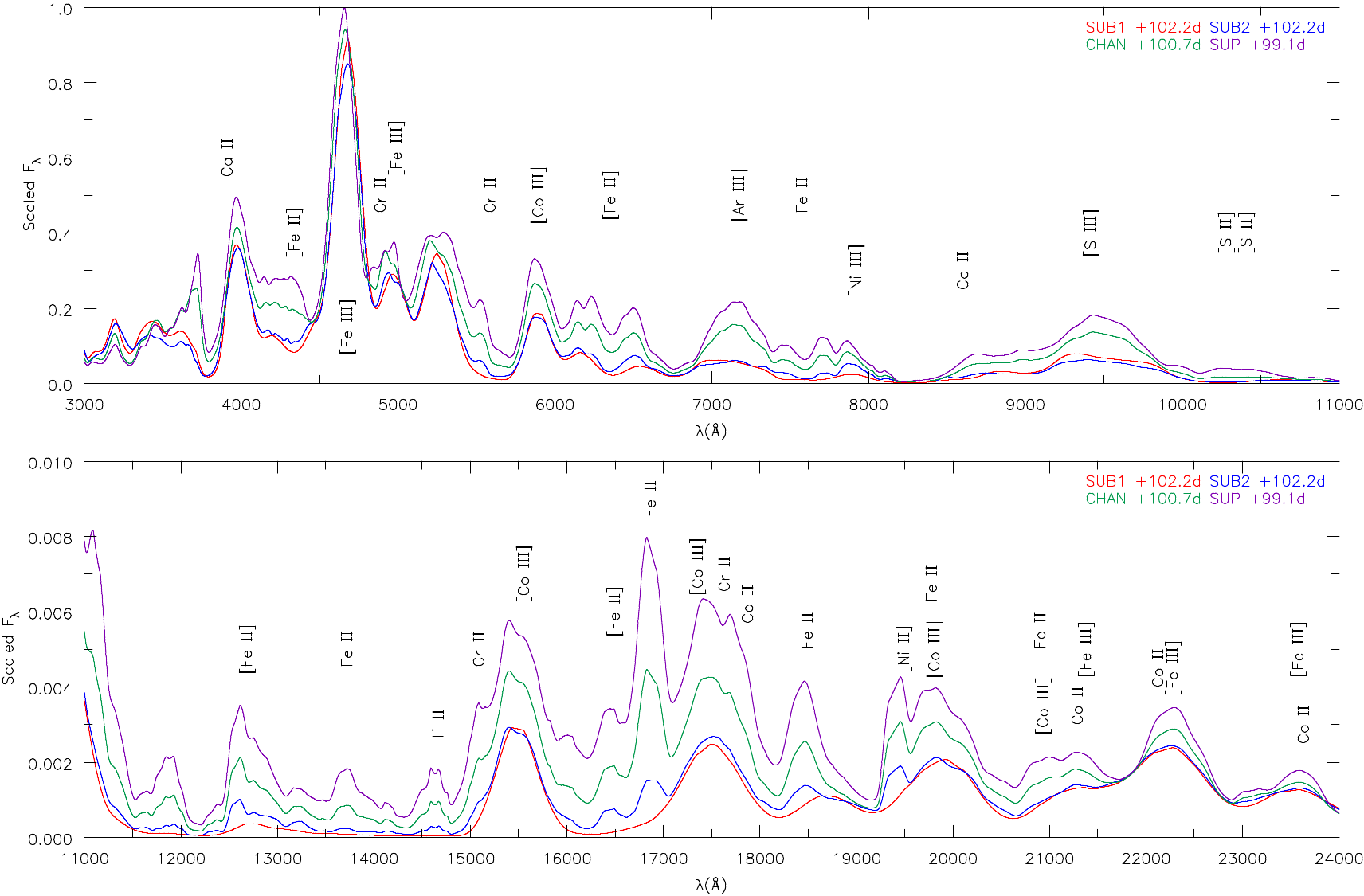}
\end{minipage}
\caption{Spectral comparison between models in the nebular phase at roughly $+100$ days after bolometric maximum. We have scaled all models by the same value, and plot $F_\lambda$ vs \lb. All models exhibit roughly the same amount of emission in the [\feiii] \lb4658 line, but show substantial flux differences in the NIR.  Unlike the other models, SUB1 lacks NIR \feii\ and \coii\ features. Emission lines of [\siii] \lb\lb9068,9530, and  [\ariii] \lb\lb7135,7751 are present, and increase in strength relative to [\feiii] \lb4658 as the ejecta mass, and equivalently, as the mass of IMEs, increases.}
\label{nebular_1}
\end{figure*}

\begin{figure*} 

\begin{minipage}[t]{\linewidth}
\centering
\includegraphics[scale=1]{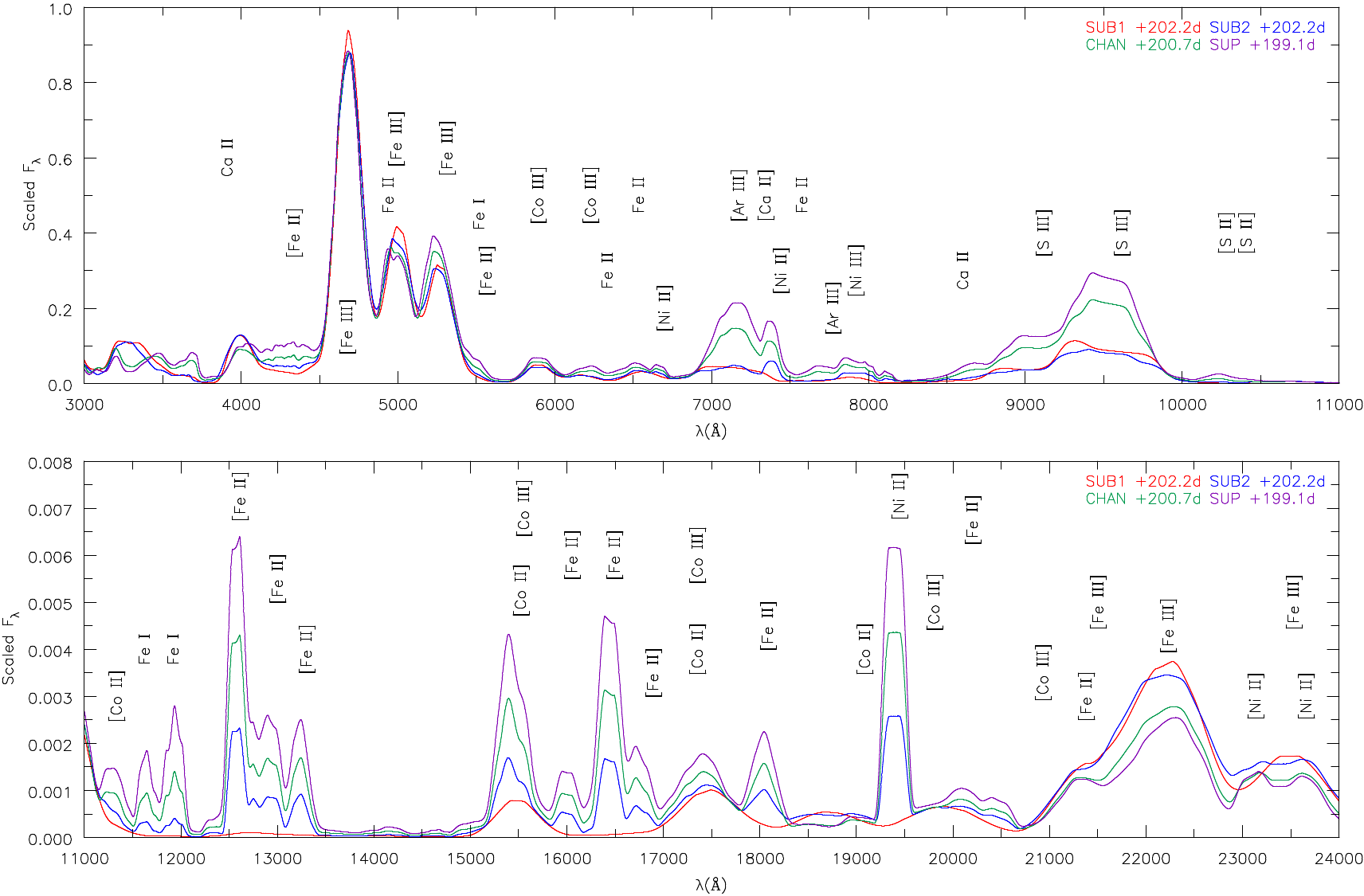}
\end{minipage}
\caption{Spectral comparison between models in the nebular phase at roughly $+200$ days after bolometric maximum. We have scaled all models by the same value, and plot $F_\lambda$ vs \lb. As at 100 days, SUB1 shows  no prominent  [\feii] or [\coii] lines. Further, spectra of SUB1 in the $J$ and $H$ bands are now substantially
different from those of SUB2. The [\siii] \lb\lb9068,9530 and the [\ariii] \lb\lb7135,7751 features have increased in strength relative to [\feiii] \lb4658. The broad emission feature covering 7000 through 7500 \AA\ is a blend of [\ariii] \lb7135, [\nkii] \lb7378, and [\caii] \lb\lb7291,7324.}
\label{nebular_2}
\end{figure*}

\skii\ \lb\lb5958,5979, observed roughly around $\sim$\lb5750, is a spectroscopic classification diagnostic for SNe Ia ($\mathcal{R}$(Si) $\equiv$ pEW(\skii\ \lb5750)/pEW(\skii\ \lb6100)) and, like other IME features, its strength correlates with ejecta mass. Table \ref{Sk2_widths} highlights the correlation of pEWs of various features with ejecta mass of our models. Post maximum (\figs~\ref{postmax_1} and \ref{postmax_2}), we see the strength of the emission increase, giving a large morphological separation between models. Calcium (as \caii) also shows the same behaviour as \skii. The absorption and emission strength of the \caii\ NIR triplet distinguishes models throughout the spectral evolution. We find that the strength of this feature correlates with ejecta mass. We further discuss the \caii\ NIR triplet and the \skii\ \lb\lb6347,6371 doublet in Section \ref{section_HVF}. 

\section{Explosion Scenario: \, SUB1 \, versus \, SUB2}%%%%%%%%%%%%%%%%%%%%%%%%%%%%
\label{section_sub1_vs_sub2}
Since the explosion process and progenitor system are unknown, we highlight and summarize useful diagnostics for distinguishing our models of the same $M_{\rm ej}$. As mentioned earlier, SUB1 comes from a detonation model of a sub-\Mch\ WD, while SUB2 comes from the DDT of a \Mch\ WD, which was scaled in density to have the same mass as SUB1 and the same \elem[56]{Ni} mass. Since SUB1 was detonated as a sub-\Mch\ WD, it had lower densities when exploded compared to SUB2 and lacks the `\elem[56]{Ni} hole'. Without the `\elem[56]{Ni} hole', SUB1 has a larger (assumed) local deposition fraction from decay positrons compared to the total decay energy deposition (\fig~\ref{edep_ratio}) for velocities less than 5000 \kms. This keeps the inner region of SUB1 hotter than SUB2, which shows stronger features of higher ionization states of IGEs as the ejecta evolves past the photospheric phase and exposes the inner iron-rich material. At nebular times, the strength of [\nkii] 1.939 $\mu$m gives a clear distinction between SUB1 and SUB2, as lower density ejecta model SUB1 does not show this feature.

Up to maximum light, SUB1 and SUB2 possess very similar spectra (\fig~\ref{premax_max_spec}), especially in the optical. However, as the photosphere begins to recede inwards differences are seen in the NIR -- \feii\ and \coii\ features are absent in SUB1 but present in SUB2 (\figs~\ref{postmax_1} and \ref{postmax_2}). Below 5000 \kms, the densities in SUB1 are roughly a factor of 3 lower than in SUB2. Further, SUB1 has a larger fraction of local radioactive heating from positrons. These factors inhibit recombination and a higher ionization persists in SUB1 compared to SUB2. The NIR region is potentially the best diagnostic for the ionization state of the ejecta in SNe Ia (shown in \figs~\ref{nebular_1} and \ref{nebular_2}).

This higher ionization, seen in post-maximum spectra (\figs~\ref{postmax_1} and \ref{postmax_2}), yields lower fluxes in the NIR. We see roughly half a magnitude difference in the post maximum $I$, $J$, and $H$ bands. However, the magnitude difference between $J$ and $H$ grows to $\sim$2 mag difference by 200 days post maximum.
%%%%%%%%%%%%%%%%%%%%%%%%%%%%%%%%%%%%%%%%%%%%%%%

\section{Additional Investigations}
In this section we present additional investigations of our ejecta models focusing on high velocity features and comparisons to observational data. We also explore shortcomings with our ejecta models.  

\subsection{High Velocity Features -- \skii\ \& \caii}

\label{section_HVF}%%%%%%%%%%%%%%%%%%%%%%%%%%%%%%%%%%%%%%%%%%%%
\begin{figure*} 
\begin{minipage}[t]{\linewidth}
\centering
\hspace{0cm}\includegraphics[scale=0.80]{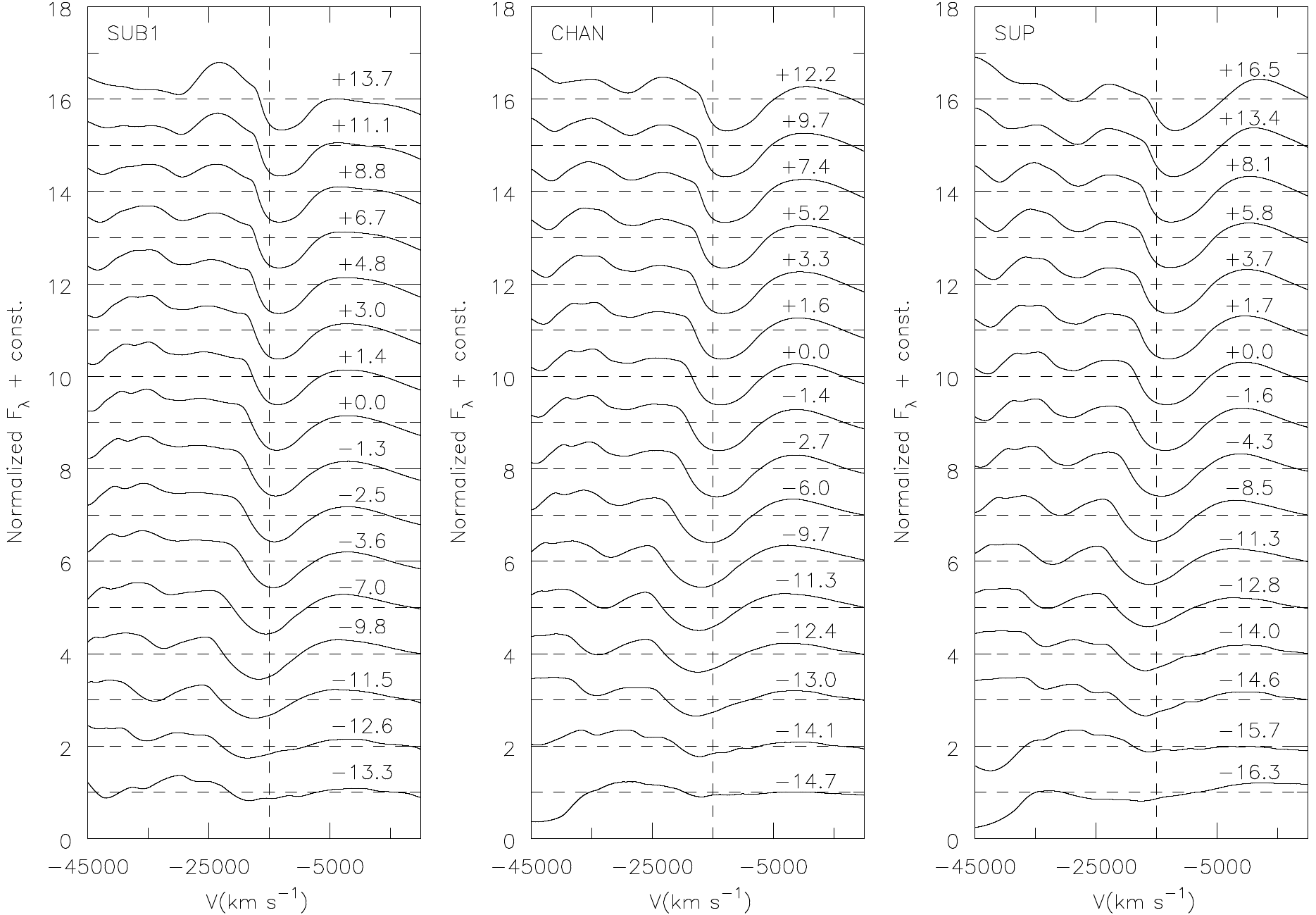}
\end{minipage}
\caption{Normalized flux (according to Eqn. \ref{normalize} between 5800-6500 \AA) of the \skii\ \lb6347,6371 doublet relative to bolometric maximum plotted in velocity space shifted relative to \lb6355. Note the lack of \skii\ HVFs. A vertical line at $-15\,000$~\kms\ is included as a reference.}
\label{HVF_Sk2}
\end{figure*}
\begin{figure*} 
\begin{minipage}[t]{\linewidth}
\centering
\hspace{0cm}\includegraphics[scale=0.80]{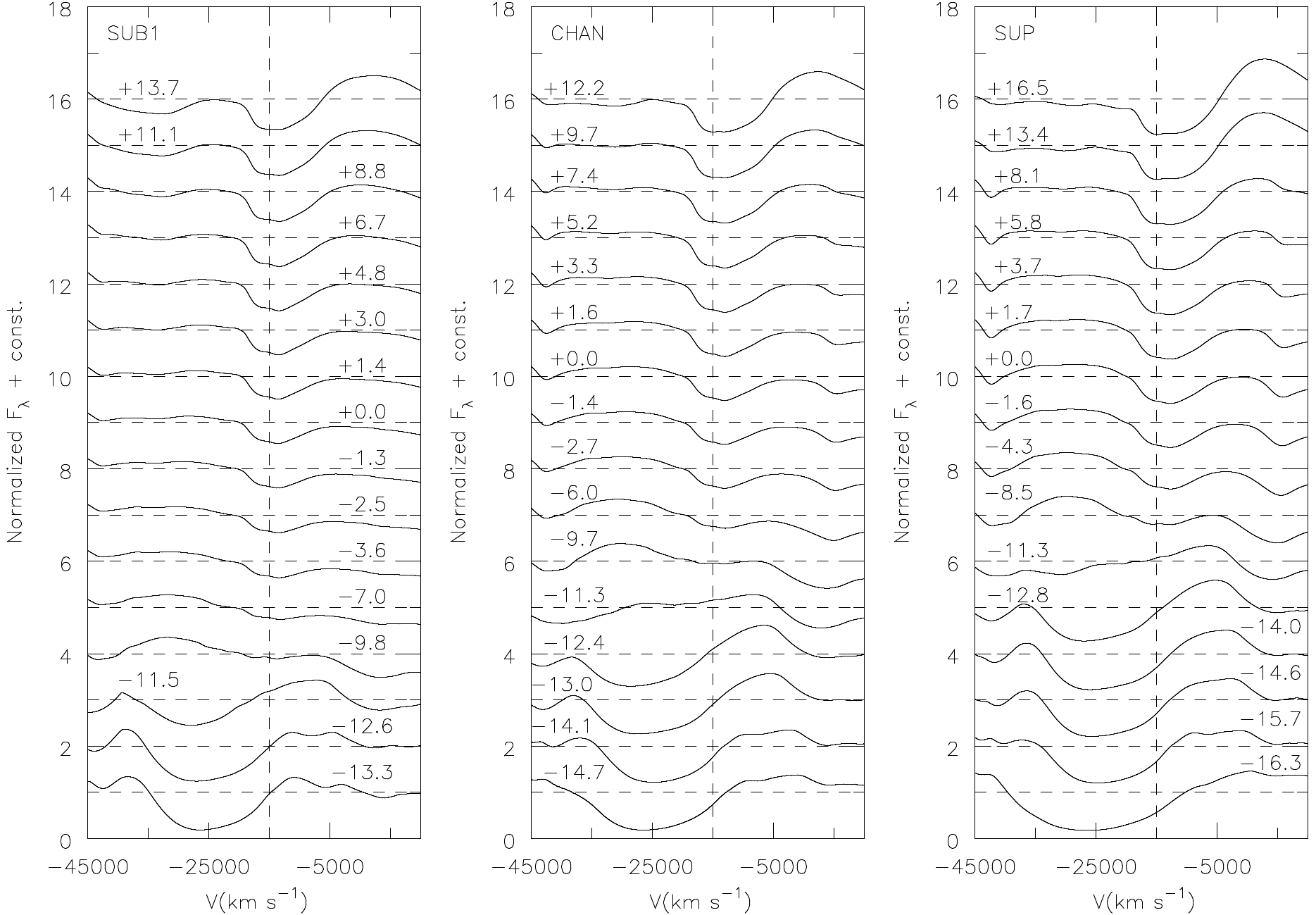}
\end{minipage}
\caption{Normalized flux (according to Eqn. \ref{normalize} between 7000-9000 \AA) of the \caii\ triplet relative to bolometric maximum plotted in velocity space shifted relative to \lb8662. Note prior to $\lesssim$-11 days, HVFs are present without a photospheric component. After the HVF disappears, the photospheric component becomes visible. A vertical line at $-15\,000$~\kms\ is included as a reference.}
\label{HVF_Ca2}
\end{figure*}

High velocity features (HVFs) are absorption features, seen in the strongest lines, that show a distinct difference in velocity (often early and prior to maximum), by more than a few thousand \kms\, from the lower velocity, photospheric component \citep{Gerardy2004,Mazzali2005a,Mazzali2005b}. Note the two strongest components of the \caii\ NIR triplet (\lb8542 \& \lb8662) are separated by $\sim$4000\,\kms. Thus, any single \caii\ NIR profile may show an absorption feature with two components separated by a few thousand \kms\, which is different from a HVF. In many SNe Ia, HVFs have even been observed at maximum for the \caii\ NIR triplet but not for \skii\ \lb\lb6347,6371 \citep{Childress2014}. There is no clear indication when HVFs start to disappear in all observed cases.  \cite{Silverman2015} state that the HVF \caii\ triplet begins to disappear around $-1$ days prior to maximum for $\Delta M_{15}(B)=1.4$ to 1.6 mag, however discoveries of HVFs are potentially biased towards those that persist closer to maximum light. 

Shown in \figs~\ref{HVF_Sk2} and \ref{HVF_Ca2} is the evolution of the \skii\ doublet and the \caii\ NIR triplet, with a vertical line at $-15\,000$\,\kms\ as a reference. In all models, \caii\ HVFs are seen before bolometric maximum ($\lesssim-11$ days). However, no HVF for \skii\ \lb\lb6347,6371 is seen. Notice the striking difference in the  pre-maximum \caii\ triplet profile ($\lesssim -11$ days) and the profile at later dates.
The lack of a \skii\ doublet HVF could just be a byproduct of atomic physics. Although  both the \caii\ NIR triplet and the \skii\ doublet are not resonance transitions, the lower level of the \caii\ triplet is metastable. The lower level of the \skii\ \lb6355 doublet is the 4s state which is coupled to the ground state by a permitted transition. Therefore, when compared to the \skii\ \lb6355 doublet, the \caii\ NIR triplet persists longer because the metastable lower level population persists longer. 

\begin{figure} 
\includegraphics[scale=0.8]{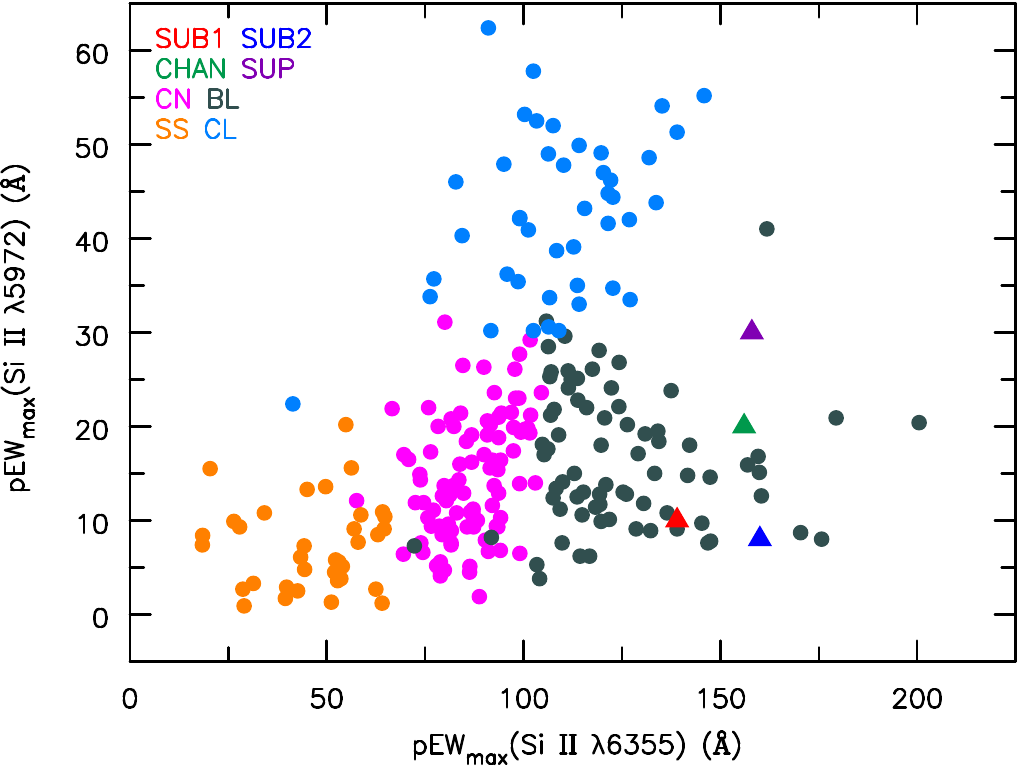}
\caption[]{Plot of the pEWs of the \skii\ $\lambda$6355 and $\lambda$5972 features along with the data from \cite{Blondin2012}. CN, BL, SS, and CL correspond to ``core normal", ``broad line", ``shallow silicon", and ``cool" classifications defined by \cite{Branch2006}. Our models lie clustered near the BL classification.}
\label{BranchTypes}
\end{figure}

By defining a straight line between the maxima on either side of the absorption profiles of these \skii\ features, we are able to compute our models' pEWs (listed in Table \ref{Sk2_widths}). Comparing our work to fig.~8 of \cite{Blondin2012}, we find our spectra are clustered around those labelled broad-lined Ia as seen in \fig~\ref{BranchTypes}. \cite{Branch2006} looked at the pEWs of \skii\ features near $\lambda 6100$ and $\lambda 5750$ in these spectra at maximum in order to group these spectra in different classifications: ``core-normal", ``broad-line", ``shallow-silicon", and ``cool". For the most massive model (SUP), it might fall under the ``cool" classification from \cite{Branch2006}, but it lacks the strong \tkii\ absorption.

\begin{table}
\centering
 \begin{tabular}{| l  | c | c | c | c |} 
 \hline
 Model & \skii\ \lb5750 & \skii\ \lb6100 & \oi\ \lb7400 & \mgii\ \lb8700 \\ [0.5ex] 
 \hline\hline
 SUB1 & 10 & 139 & 17 & $\lesssim$ 1 \\
 SUB2 & 8 & 160 & 9 & $\sim$ 1 \\ 
 CHAN & 20 & 156 & 32 & 32 \\
 SUP & 30 & 158 & 50 & 53 \\
 \hline
 \end{tabular}
\caption[]{Approximate pEW (\AA) based on a straight line across the profile of the absorption feature.}
\label{Sk2_widths}
\end{table}

\subsection{Comparison to Data}%%%%%%%%%%%%%%%%%%%%%%%%%%%%%%%%%%
\label{section_data_compare}
Here we present both light curve and spectral comparisons to data for a span of spectral epochs. We focus on SNe Ia that have claimed \elem[56]{Ni} masses similar to that of our models (0.6 \Msun) or similar \delm\ and those tagged as sub-\Mch\ (SN2005el), \Mch\ (SN1995D), and similar Branch types (SN2001ay). We used the supernova identification program SNID \citep{Blondin2007} on models at bolometric maximum to find additional SNe Ia to compare (SN1994ae). The spectra are taken from the CfA Supernova Archive \citep{Blondin2012}. Archived light curve photometry is taken from Open Supernova Catalog \citep{Guillochon2017}. When comparing models to observations, the spectra are normalized between $\lambda_{\rm min}$=4000 \AA\ and $\lambda_{\rm max}$=7000 \AA, such that
\begin{equation}\label{normalize}
	\frac{1}{\lambda_{\rm max}-\lambda_{\rm min}}\int^{\lambda_{\rm max}}_{\lambda_{\rm min}}F_\lambda d\lambda=1 \text{ erg cm$^{-2}$ s$^{-1}$ \AA$^{-1}$}
\end{equation}
Normalizing spectra allows us to better compare spectral features, removes uncertainties in distance, and compensates for small differences in \isoni\ mass. To compare LCs, we correct for extinction using the CCM reddening law \citep{CCM1989} and literature $E(B-V)$ and $R_V$ values. We normalize the LCs by adding a constant offset (model and object dependent), such that $B_{\rm max}$ = 0 mag at $t(B_{\rm max})$. We also shift the LCs so that time of $B$-band maxima agree. Thus, uncertainties in distance and explosion time are reduced. A constant value of 0.05 mag is included with the photometric error bars for uncertainty in reddening. $K$-corrections, expected to be small, have not been applied.  Photometric band magnitudes and bolometric luminosities of the models at maximum are provided in Table~\ref{lc_data}.

\subsubsection{SN1994ae}%%%%%%%%%%%%%%%%%%%%
\begin{figure*} 
\begin{minipage}[t]{\linewidth}
\includegraphics[scale=0.9]{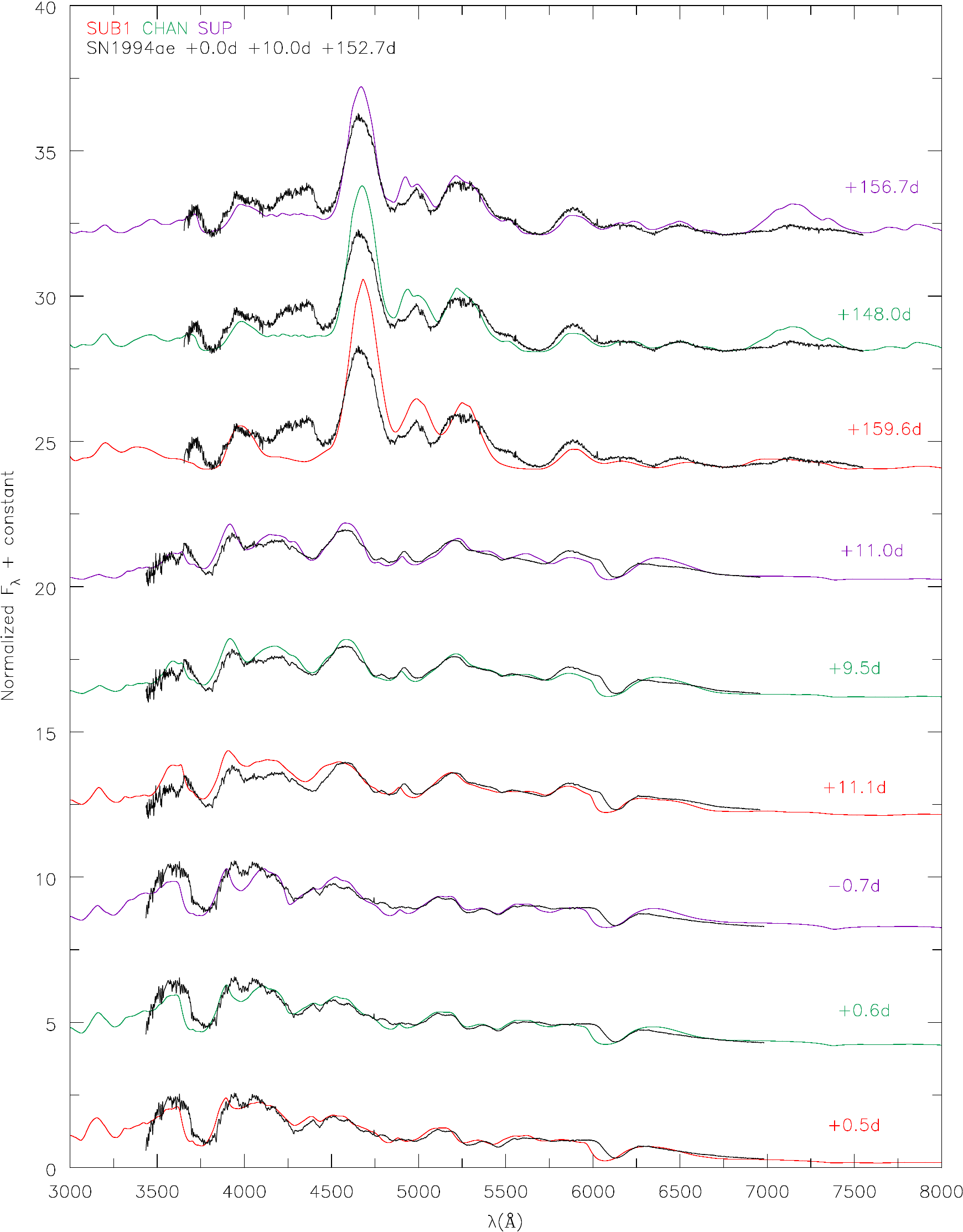}
\end{minipage}
\caption[]{SN1994ae compared to our models at epochs $+0.0$, $+10.0$, and $+152.7$ days relative to $B$-band maximum, which occurred on MJD 49685.5. For comparison, we reddened our models using $E(B-V)=0.0226$ mag and $R_V=3.1$ \citep{Jha2007}. All fluxes have been normalized between 4000-7000 \AA\ according to Eqn. \ref{normalize}. The observational data was taken from public CfA data \url{https://www.cfa.harvard.edu/supernova/SNarchive.html}. At $+0$ days, we see that our model spectra do not reproduce the velocity of the \skii\ \lb6347,6371 doublet and the UV \skii\ triplet. The spectra show evidence of the \skiii\ triplet (\lb\lb4553,4568,4575) absorption as do our SUB models (an indication of high ionization). Spectra at later epochs indicate cooler ejecta and model SUP is closest to reproducing the features. However, at nebular times (+152.7 days) our model optical \feiii\ lines appear too strong and the [\siii] \lb\lb9068,9530 lines appear to be absent in observed spectra.}
\label{SN1994ae_spec}
\end{figure*}

\begin{figure*} 
\begin{minipage}[t]{\linewidth}
\hspace*{-1.0cm}
\includegraphics[scale=0.85]{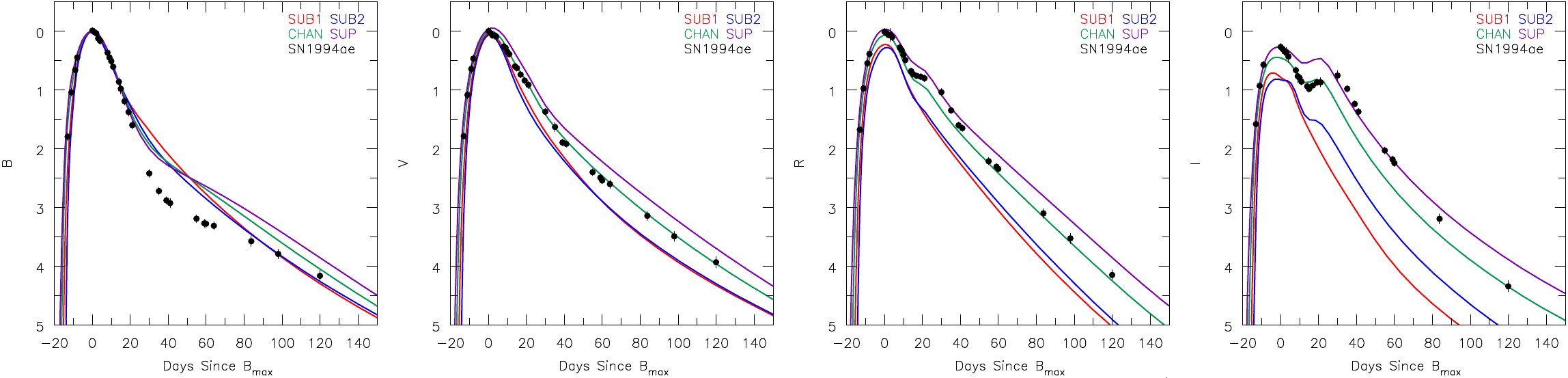}
\end{minipage}
\caption[]{SN1994ae multi-band LCs normalized to $B_{\rm max}$ and corrected for reddening (see Section \ref{SN1994ae}). Our $B$-band LC agrees until 20 days post maximum. Afterwards, our LCs show roughly half a magnitude more flux. The $V/R$-band LC agrees with our model CHAN. Our models disagree with observations in the location of the $I$-band's second peak and more so for models with lower ejecta mass. Archived light curve photometry is taken from Open Supernova Catalog \citep{Guillochon2017}, with references from the CfA Supernova Archive (\url{https://www.cfa.harvard.edu/supernova/SNarchive.html}) and \cite{Riess2005}.}% the Sternberg Astronomical Institute Supernova Light Curve Catalogue (\url{http://dau.itep.ru/sn/node/72}), \cite{Tsvetkov1997}, \cite{Riess1999CFA1},  \cite{Ho2001}, \cite{Altavilla2004}, and \cite{Riess2005}.}
\label{SN1994ae_LC}
\end{figure*}

\label{SN1994ae}
SN1994ae exploded in NGC 3370 \citep[$z=0.0043$ --][]{Riess1999CFA1,Jha2007} and was first discovered on 14 November 1994 by \cite{vanDyk1994}. It reached $B$-band maximum ($m_B=13.21$ mag) on MJD 49685.5 with  \delm=0.96 mag \citep{Riess1999CFA1,Jha2007}. For comparison, we reddened our models using $E(B-V)=0.0226$ mag and $R_V=3.1$ \citep{Jha2007}. \fig~\ref{SN1994ae_spec} shows the spectral comparison of SN1994ae at $+0.0$, $+10.0$, and $+152.7$ days after $B$-band maximum and normalized LCs relative to band maximum are shown in \fig~\ref{SN1994ae_LC}.

At $+0$ days, our model spectra do not reproduce the velocity of the \skii\ \lb6355 doublet and UV \skii\ triplet. Results by \cite{Dessart2014a} suggest SNe Ia resulting from pulsational-delayed detonations (PDD) retain more unburnt carbon and have little mass at high velocity ($\gtrsim$15\,000 \kms) due to pulsations. Therefore, spectral features of SN1994ae might be best explained by PDD modeling, and would resemble similar radiative properties of DDT models. 
The spectra also show evidence of the \skiii\ triplet (\lb\lb4553,4568,4575) absorption as in our SUB models (an indication of high ionization). Later spectra show cooler ejecta and model SUP is closest to reproducing the features. However, at nebular times (+152.7 d) our model optical [\feiii] lines appear too strong and [\siii] \lb\lb9068,9530 are absent in the observational data. 

For the light curve comparison, we shifted the LCs  to give the same time of $B_{\rm max}$ and reddened the models with $A_B$ = 0.091, $A_V$ = 0.070, $A_R$ = 0.057, and $A_I$ = 0.041 mag, obtained using $E(B-V)=0.0226$ mag and $R_V=3.1$ from \cite{Jha2007}. We normalized the light curves to 0 mag at $B_{\rm max}$ and shifted the observational data by 12.98 mag. We see in \fig~\ref{SN1994ae_LC} that our $B$-band LC is consistent until 20 days post maximum, where our LCs begin showing roughly half a magnitude more flux. Model CHAN matches well the $V/R$-band observations. However, our models fail to reproduce the second peak in the $I$-band, and the disagreement is greater in lower mass models.


\subsubsection{SN1995D}%%%%%%%%%%%%%%%%%%%%
\label{SN1995D}
SN1995D exploded in NGC 2962 and was discovered on 10 February 1995 \citep{Nakano1995}. Its redshift is $z=0.0067$, and it reached $B$-band maximum ($m_B$=13.44 mag) on MJD 49768.7 \citep{Riess1999CFA1,Jha2007}. SN1995D has been argued as having a \elem[56]{Ni} mass of about 0.58 \Msun\ and an ejecta mass around 1.45 \Msun\ \citep{Childress2015}. For comparison, we reddened our models using $E(B-V)=0.026$ mag and $R_V=3.1$ \citep{Jha2007}.

\begin{figure*} 
\begin{minipage}[t]{\linewidth}
\includegraphics[scale=0.9]{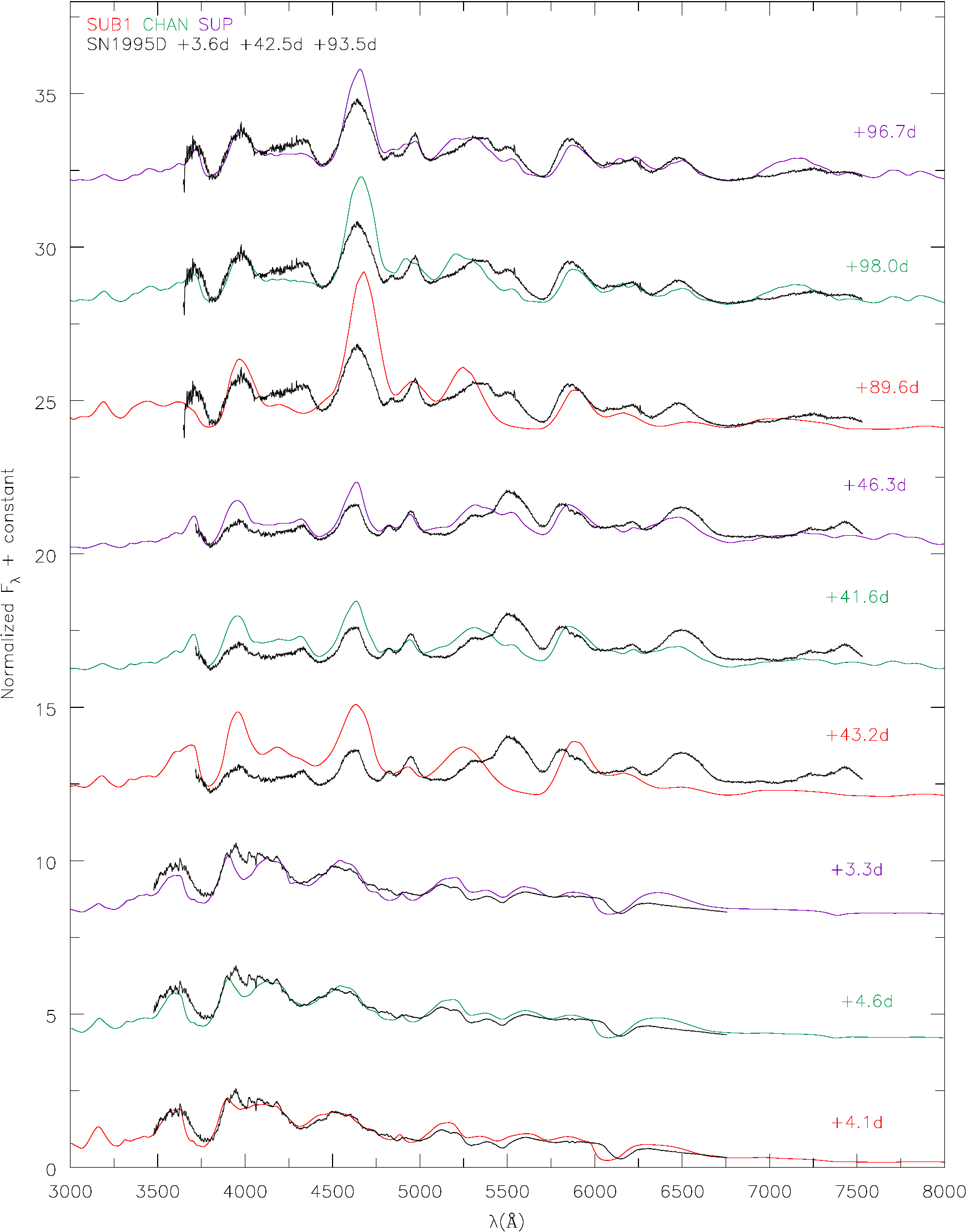}
\end{minipage}
\caption[]{SN1995D compared to models at times $+3.6$, $+42.5$ and $+93.5$ days relative to $B$-band maximum which occurred on MJD 49768.7. We have corrected for redshift with a value of 0.0067. We reddened the models with $E(B-V)$=0.026 mag and $R_V$=3.1. All fluxes have been normalized between 4000-7000 \AA\ according to Eqn. \ref{normalize}. The observational data was taken from public CfA data \url{https://www.cfa.harvard.edu/supernova/SNarchive.html}. $+3.6$ days shows good qualitative agreement with SUB1, except our models have the \skii\ \lb6355 doublet formed at a higher velocity. SUB1 also matches the UV spectrum shortward of 4000 \AA. Later epochs ($+42.5$ and $+93.5$ days) show a better agreement to model SUP and to CHAN, due to lower temperatures and ionization. Roughly all features at $+93.5$ days are matched by SUP.}
\label{SN1995D_spec}
\end{figure*}

\fig~\ref{SN1995D_spec} shows the spectral comparison of SN1995D at $+3.6$, $+42.5$, and $+93.5$ days after $B$-band maximum. The early epochs ($+3.6$ days) show good qualitative agreement with SUB1 except our model shows a larger blueshifted \skii\ doublet. This may be best explained by a PDD model \citep{Dessart2014a}. SUB1 also matches the UV spectrum shortward of 4000 \AA. At later epochs like $+42.5$ and $+93.5$ days, we see a better agreement to model SUP and to CHAN, due to lower temperatures and ionization. Roughly all features at $+93.5$ days are matched by SUP. 
Despite the calculated \elem[56]{Ni} and ejecta mass being closest to CHAN, SN1995D shows only moderate qualitative agreement at later epochs. SN1995D transitions from looking like our SUB1 into that of SUP from early to late epochs. 

To compare light curves, we shifted the LCs to give the same time of $B_{\rm max}$ and  reddened the models with $A_B$ = 0.106, $A_V$ = 0.081, $A_R$ = 0.066, and $A_I$ = 0.048 mag, obtained using $E(B-V)=0.026$ mag and $R_V=3.1$ \citep{Jha2007}. We normalized the light curves to 0 mag at $B_{\rm max}$ and shift the observational data by 13.35 mag. 

\begin{figure*} 
\begin{minipage}[t]{\linewidth}
\hspace*{-1.0cm}
\includegraphics[scale=0.85]{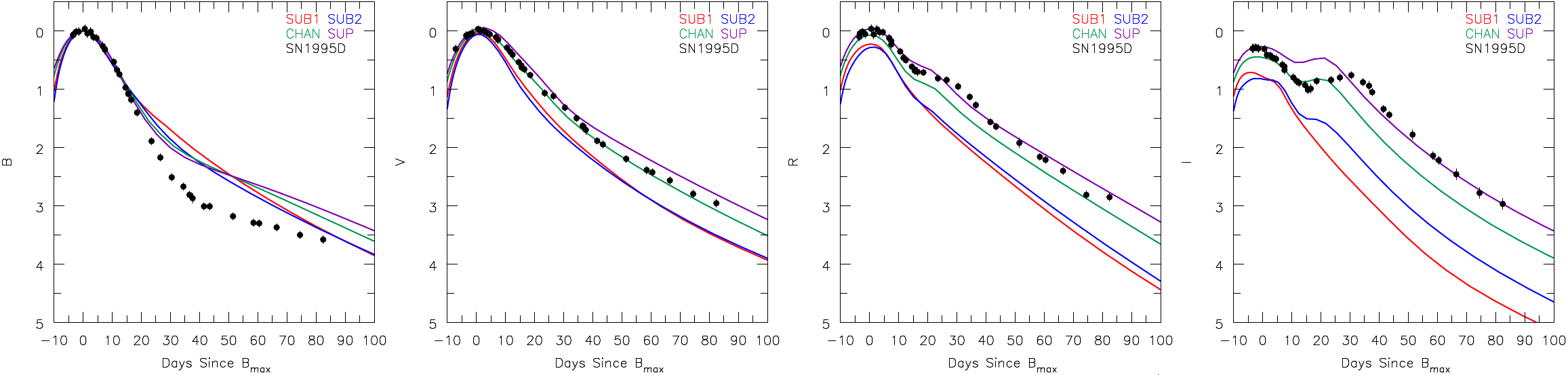}
\end{minipage}
\caption[]{SN1995D multi-band LCs normalized to $B_{\rm max}$ and corrected for reddening (see Section \ref{SN1995D}). $B$-band LC agrees until $\sim$12 days post maximum. Afterwards, our LCs show roughly half a magnitude more flux. The $V/R$-band LCs qualitatively agree with our CHAN model although our models fail to reproduce the second peak in the $I$-band. However, model SUP agrees with the observational data at late times. Archived light curve photometry is taken from Open Supernova Catalog \citep{Guillochon2017}, with references from CfA Supernova Archive \url{https://www.cfa.harvard.edu/supernova/SNarchive.html} and \cite{Riess1999CFA1}.}
\label{SN1995D_LC}
\end{figure*}

In \fig~\ref{SN1995D_LC}, we see that our $B$-band LC is consistent with all models until $\sim$12 days post maximum, where our LCs then begin showing roughly half a magnitude more flux. Model CHAN matches well the $V/R$-band observations. Our models fail to reproduce the second peak in the $I$-band (it occurs 10 to 20 days too early) although the SUP model matches the data at late times. Model SUP also seems consistent with the peak flux ratios in SN1995D.

\subsubsection{SN2001ay}%%%%%%%%%%%%%%%%%%%%
\label{SN2001ay}
SN2001ay exploded outside IC 4423 and was discovered on 18 April 2001 by \cite{SwiftLi2001}. \cite{Krisciunas2011} and references therein cite its redshift as $z=0.0302$ and indicate that it reached $B$-band maximum ($M_B=-19.19$ mag) on 23 April 2001. For spectral comparison, we reddened our model spectra using $E(B-V)_{MW}$ = 0.026 mag, $E(B-V)_{\rm host}$ = 0.072 mag, and $R_V$ = 3.1 \citep{Krisciunas2011}. \cite{Krisciunas2011} states a $M(^{56}\text{Ni})$ of (0.58$\pm$0.15)/$\alpha$ \Msun, for an $\alpha=L_{\rm max}/E_{\rm Ni}$, typically between 1--1.2. 
%Given the close proximity the models have to SN2001ay in Fig.~\ref{BranchTypes} and the comparable $M_{^{56}\text{Ni}}$, we explore the spectral similarities.
Given the close proximity between the estimated \elem[56]{Ni} mass for SN2001ay and that of our model set, we explore the spectral similarities. 

\begin{figure*} 
\begin{minipage}[t]{\linewidth}
\includegraphics[scale=0.9]{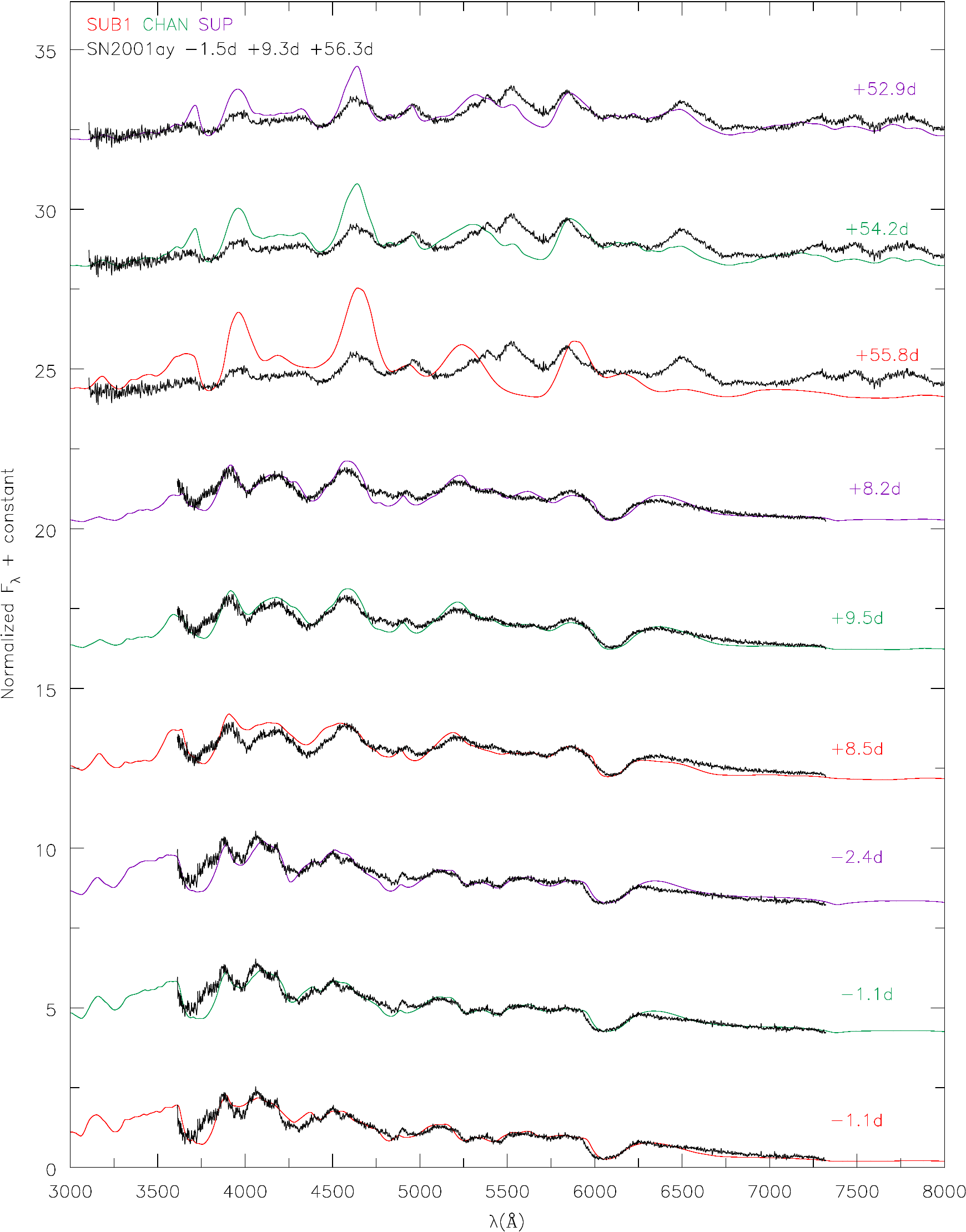}
\end{minipage}
\caption[]{SN2001ay compared to models at epochs $-1.5$, $+9.3$, and $+56.3$ days relative to $B$-band maximum. Observed spectra have been corrected for a redshift of $z=0.0302$. Models are reddened using $E(B-V)_{MW}$=0.026 mag, $E(B-V)_{\rm host}$=0.072 mag, and $R_V$=3.1. All fluxes are normalized in the range of 4000-7000 \AA\ according to Eqn. \ref{normalize}. Observational data was taken from public CfA data \url{https://www.cfa.harvard.edu/supernova/SNarchive.html}. Early on, models reproduce velocity of the \skii\ \lb\lb6347,6371 doublet. Models show stronger \skii\ \lb\lb5041,5056,5056.3 triplet absorption seen around 4800 \AA, and also show discrepancies in \feii\ \lb5018, seen around 4900 \AA. SN2001ay shows similar \skiii\ \lb\lb4553,4568,4575 triplet absorption as CHAN. No model reproduces the shape of the UV feature, most likely a discrepancy of formation velocity for the \caii\ H\&K lines. Later spectral phases ($+9.3$ and $+56$ days) are best reproduced by SUP, indicating much lower temperature towards the innermost ejecta. SUP, however, shows a stronger absorption feature of the \caii\ NIR triplet. Given the discrepancies we see at $-1.5$ days with the \caii\ H\&K lines, SN2001ay shows a lack of strong \caii\ features.}
\label{SN2001ay_spec}
\end{figure*}

\fig~\ref{SN2001ay_spec} shows the spectral comparison for epochs $-1.5$, $+9.3$, and $+56.3$ days relative to $B$-band maximum. All models provide a good qualitative fit to the optical spectrum at $-1.5$ days, with SUP exhibiting the worst fit. While all models fit the \skii\ \lb\lb6347,6371 doublet in absorption strength and velocity, our models show stronger absorption in the \skii\ \lb\lb5041,5056,5056.3 triplet around 4800 \AA. Blended with this feature
is absorption arising from \feii\ \lb5018, and this is also somewhat too strong in the models. The biggest discrepancy between model and observation for the blend occurs for model SUP. SUB1 lacks absorption at $\sim$4000 \AA, which is clearly present in the observations, and all of the other models. No model reproduces the shape of the UV absorption near 3700 \AA, which could be due to a discrepancy with the \caii\ H\&K lines.  

Later, model SUP qualitatively agrees the SN2001ay spectra at $+9.3$ and $+56.3$ days best. At $+9.3$ days, SUP shows agreement despite its stronger \feii\ absorption lines around 4800 \AA. At $+56.3$ days, the spectra is dominated by \feii\ features, which SUP matches well given its cooler temperatures and lower ionization. Models SUB1, SUB2, and CHAN are too highly ionized, and exhibit too much emission from higher ionization states such as \feiii. Despite matching much of the optical spectrum, SUP does not match well the absorption features associated with the \caii\ NIR triplet and the \caii\ H\&K lines. Given the discrepancy with calcium at $-1.5$ days, this may indicate that the calcium abundance is too high, or that the distribution in velocity space is incorrect.

\begin{figure*} 
\hspace*{-1.0cm}
\includegraphics[scale=0.85]{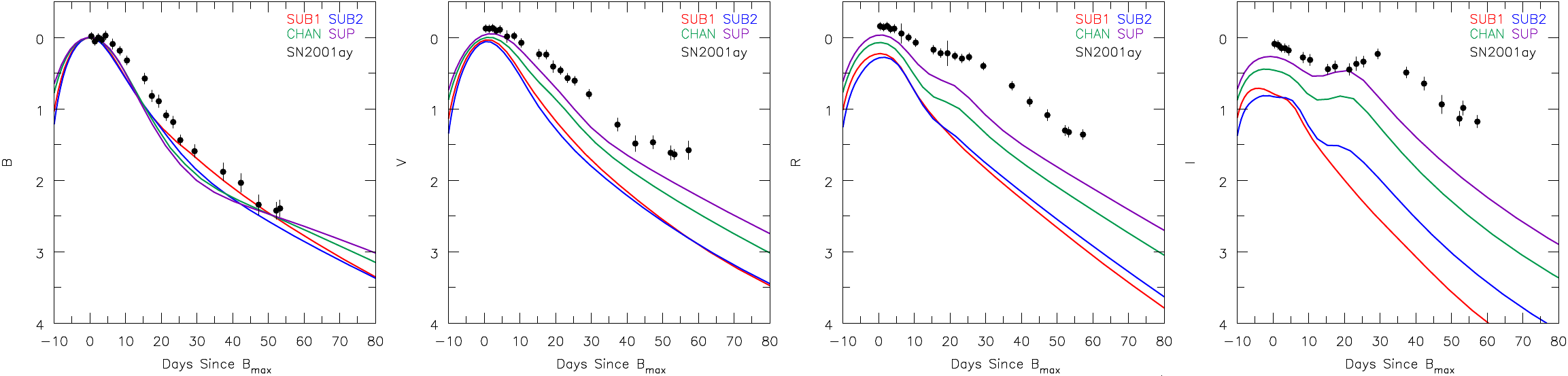}
\caption[]{Multi-band LCs for SN2001ay normalized to $B_{\rm max}$ and corrected for reddening (see Section \ref{SN2001ay}).  Our models fail to reproduce the post maximum decline except for $U/B$-bands. Models show too little $V/R/I$ flux in the decline post maximum. However, the peak flux ratio between bands matches to model SUP. A larger reddening and an adjustment of the time of $B$-band maximum would improve the comparison. Archived light curve photometry is taken from Open Supernova Catalog \citep{Guillochon2017}, with references from \cite{Ganeshalingam2010}, and \cite{Silverman2012}.}
\label{SN2001ay_LC}
\end{figure*}

For the  light curve comparison, we reddened the models with $A_B$ = 0.397, $A_V$ = 0.307, $A_R$ = 0.148, and $A_I$ = 0.178 mag, by combining host and MW values as $E(B-V)$=0.098 mag and $R_V$=3.1 \citep{Krisciunas2011}. We  normalize the light curves to 0 mag at $B_{\rm max}$ and shift the observed data by 16.35 mag. \fig~\ref{SN2001ay_LC} shows our models fail to reproduce the post maximum decline except for $U/B$-bands. Our models show too little $V/R/I$ flux in the decline post maximum, but SUP agrees in peak flux ratios between bands with SN2001ay.

\subsubsection{SN2005el}%%%%%%%%%%%%%%%%%%%%
\label{SN2005el}
SN2005el exploded in NGC 1819 and was discovered on 19 September 2005 \citep{Madison2005} at a redshift of $z=0.0148$ \citep{Hicken2009}. It reached $B$-band maximum ($m_B=14.84$ mag) on MJD 53646.4 \citep{Hicken2009}. \cite{Scalzo2014a} classified SN2005el as having 0.9 \Msun\ of ejecta as well as 0.54 \Msun\ of \elem[56]{Ni}, which, considering the errors in the determinations, are close to our models SUB1 and SUB2. For comparison, we applied reddening to our models using $E(B-V)=0.136$ mag and $R_V=3.1$ (an $E(B-V)$ value that is higher than that stated in \cite{Scalzo2014a}).

\begin{figure*} 
\begin{minipage}[t]{\linewidth}
\includegraphics[scale=0.9]{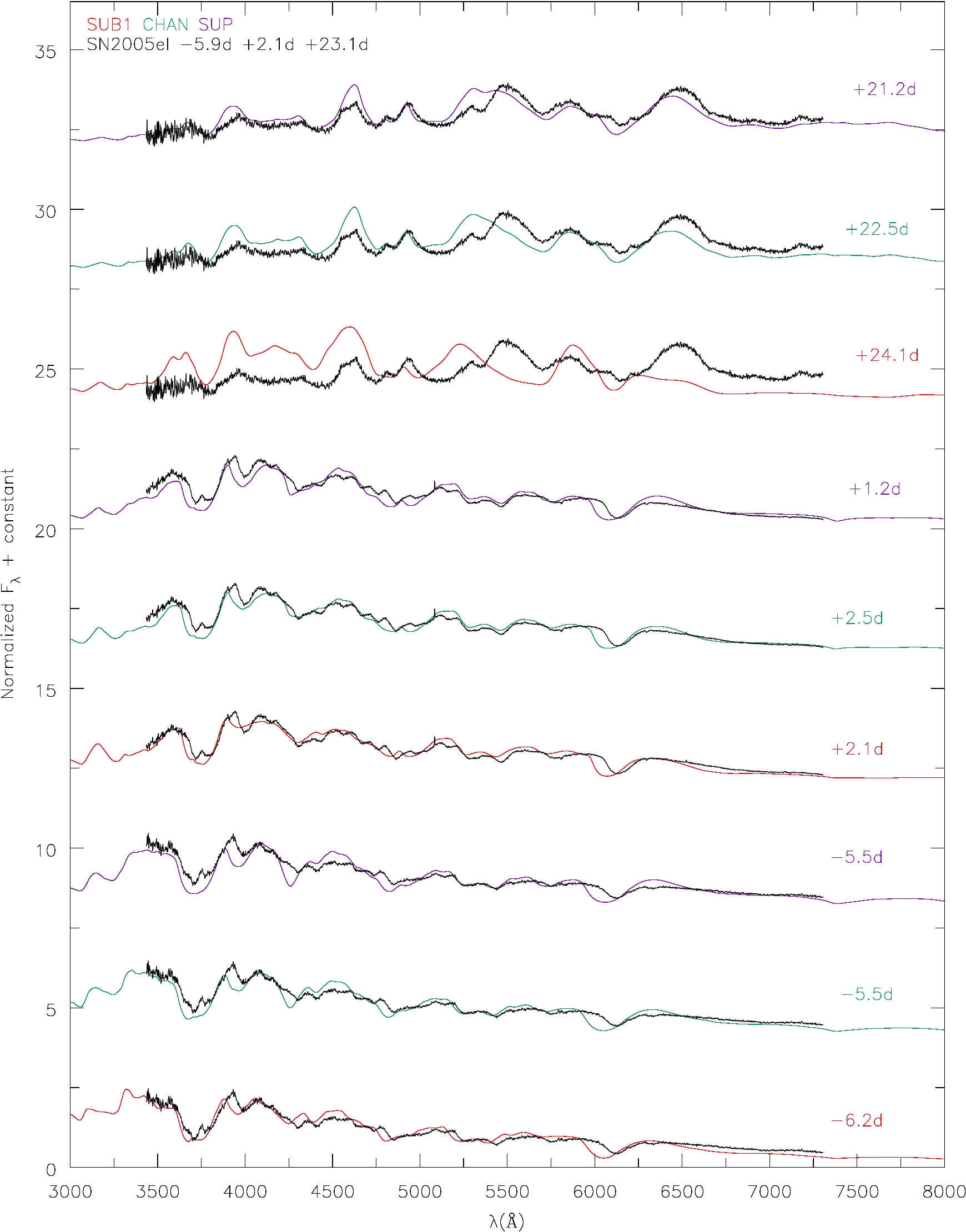}
\end{minipage}
\caption[]{SN2005el compared to models at times $-5.9$, $-4.9$, and $+23.1$ days relative to $B$-band maximum. Observed spectrum is corrected for a redshift of 0.0148. Models are reddened with $E(B-V)$=0.136 mag and $R_V$=3.1. All fluxes are normalized between 4000-7000 \AA\ according to Eqn. \ref{normalize}. This observational data was taken from public CFA data \url{https://www.cfa.harvard.edu/supernova/SNarchive.html}. Around $-5$ days, our models do not reproduce the \skii\ \lb\lb6347,6371 doublet. Models show line formation at higher velocities. Unlike the \skii\ doublet, CHAN and SUP do reproduce the \sii\ ``w" feature. UV spectra is poorly reproduced due to formation velocities of \skii\ and \caii\ H\&K features in this region. If the \caii\ H\&K lines are formed a few thousand \kms\ away from the \skii\ \lb\lb3854,3856,3862 triplet (which our models show both sum to the broad absorption feature around 3700 \AA), it is likely to result in the spike seen at the bottom of the 3700 \AA\ absorption feature. SN2005el also shows a strong absorption profile around 4400 \AA, which we suggest to be the \skiii\ \lb\lb4553,4568,4575 triplet, indicating a high ionization. At $+24.9$ days, model SUP agrees qualitatively in almost all spectral features. At this epoch, SUP has the coolest inner ejecta. This agreement is surprising given the 70 per cent difference in claimed ejecta mass. The discrepancy around 5300 \AA\ could be the result of differences in the \feii\ or \crii\ absorption.}
\label{SN2005el_spec}
\end{figure*}

\fig~\ref{SN2005el_spec} shows our spectral comparison to SN2005el. The early epochs of $-5.9$ and $+2.1$ days show some qualitative agreement, mostly with SUB1 and SUB2. At this epoch, our models do not reproduce the \skii\ \lb\lb6347,6371 doublet. Our models indicate a \skii\ \lb\lb6347,6371 doublet formed at higher velocities. Therefore, spectral features of SN2005el may be best explained by a PDD model. Unlike the \skii\ doublet, models CHAN and SUP do reproduce the \sii\ `w' feature. Since our models show a higher blue-shifted \skii\ doublet, it is not surprising that our UV does not match, given other \skii\ and \caii\ H\&K features in this region. 
%If the \caii\ H\&K lines are shifted in velocity space relative to the \skii\ \lb\lb3854,3856,3862 triplet (that our models show adds to the broad absorption feature around 3700 \AA--seen in SN2008ec, for example), then it is likely to result in the spike seen at the bottom of the 3700\,\AA\ absorption feature.
If the \caii\ H\&K lines and \skii\ \lb\lb3854,3856,3862 triplet are separated by thousands of \kms, then it is likely to result in the spike seen at the bottom of the 3700\,\AA\ absorption feature, whereas our models show one broad absorption feature around 3700 \AA\ -- seen in SN2008ec, for example).
 Given the strong absorption profile around 4400 \AA, we suggest this is the \skiii\ \lb\lb4553,4568,4575 triplet, indicating a high ionization at this epoch.

At $+24.9$ days, we see that model SUP agrees qualitatively in almost all features. Other models are too blue compared to the cooler SUP model. This is surprising given the claim that SN2005el is a sub-\Mch\ SN with an ejecta mass of only 0.9 \Msun. The discrepancy around 5300 \AA\ could be the result of differences in the \feii\ or \crii\ absorption.

%One should expect the spectral evolution to resemble that of SUB1 or SUB2; however, we see that SN2005el loosely fits SUB1/SUB2 prior to maximum and does not match SUB1 or SUB2 at late epochs, matching SUP instead. There are several possible explanations for the inconsistencies. First, the \elem[56]{Ni}\ mass may be lower than 0.54\,\Msun. Second, the poor agreement in the extent of the \skii\ \lb\lb6347,6371 doublet could indicate a different explosion scenario (such as PDD mentioned earlier--little mass at high velocity).
One should expect SUB1 or SUB2 to resemble the spectral evolution of SN2005el; however, we only see that SUB1 matches prior to maximum and does not match SN2005el at late epochs, where SUP shows best agreement. There are several possible explanations for the inconsistencies. First, the \elem[56]{Ni} mass may be lower than 0.54\,\Msun. Second, the poor agreement in the extent of the \skii\ \lb\lb6347,6371 doublet could indicate a different explosion scenario (such as PDD mentioned earlier -- little mass at high velocity).

\begin{figure*} 
\includegraphics[scale=0.9]{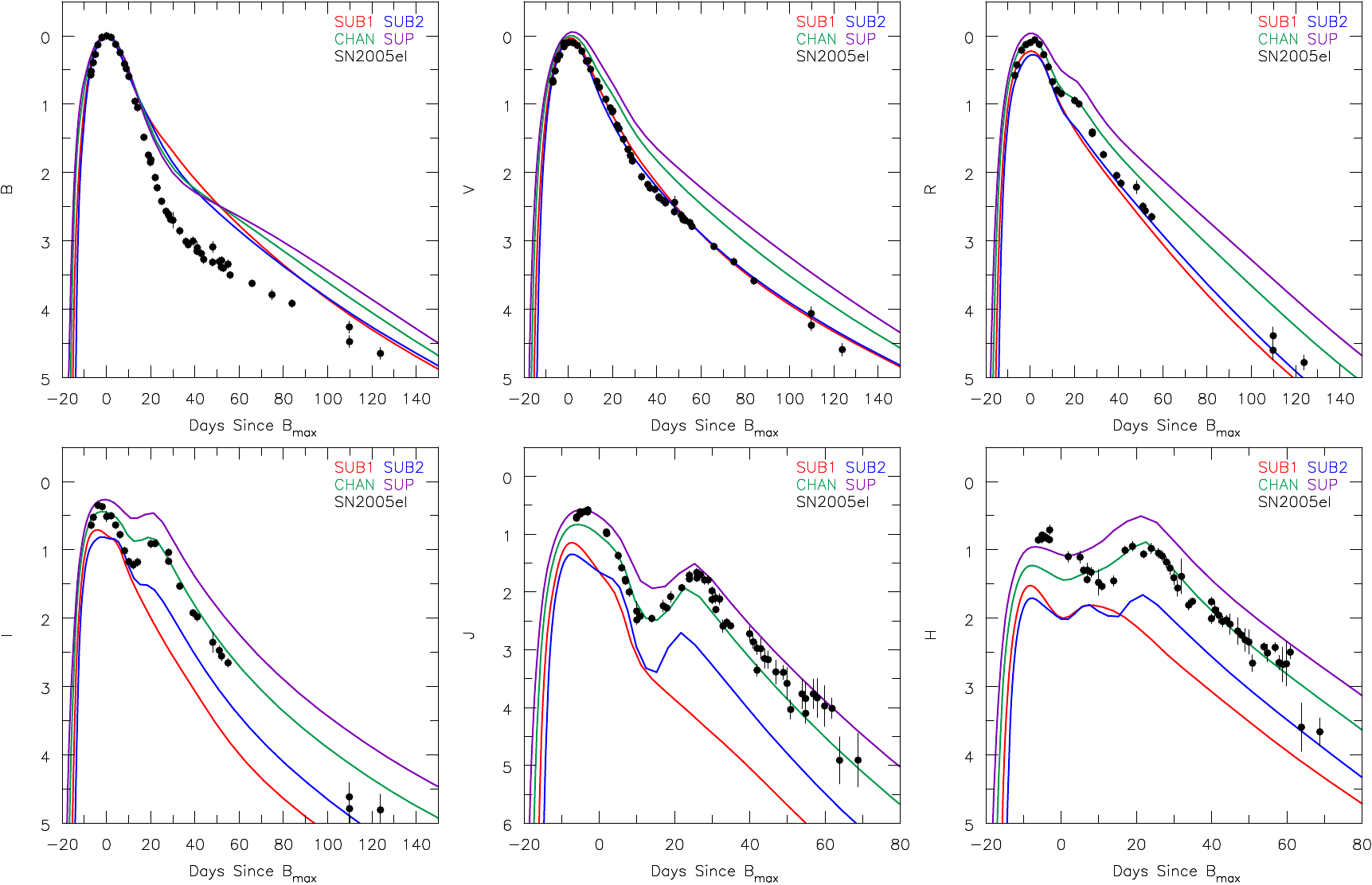}
\caption[]{Multi-band LCs corrected for reddening and normalized to $B_{\rm max}$ (see Section \ref{SN2005el}). Optical band normalized LCs of SN2005el agree in morphology with our SUB models. The NIR LCs do not agree with SUB models. Instead, the NIR LCs show better agreement with CHAN and SUP. Archived light curve photometry is taken from Open Supernova Catalog \citep{Guillochon2017}, with references from \cite{Contreras2010,Ganeshalingam2010,Silverman2012,Friedman2015}.}
\label{SN2005el_LC}
\end{figure*}

To compare light curves, we shifted the LCs relative to the time of $B_{\rm max}$ and  reddened the models with $A_B$ = 0.543, $A_V$ = 0.414, $A_R$ = 0.339, $A_I$ = 0.245, $A_J$ = 0.122, and $A_H$ = 0.077 mag, obtained using $E(B-V)=0.136$ mag and $R_V=3.1$ slightly higher than \cite{Scalzo2014a}. The light curves were normalized to 0 mag at $B_{\rm max}$ and we adjust the observational data by 14.76 mag. In \fig~\ref{SN2005el_LC}, we see the optical bands are reproduced well with our SUB models (except $B$ beyond 20 days). Although the late time behaviour in the $H$-band is reproduced, the NIR LCs do not generally agree with the SUB models. The double
peak structure in the $J$-band observations is well produced by the models, although in the $H$-band it less evident.

\subsection{Model Setbacks and Theoretical Problems} \label{section_problems}%%%%%%%%%%
When compared to observation data, our models do show a higher ionization, especially in the nebular phase. The strength of the [\feiii] \lb4658 feature is too strong compared to other optical/NIR features. Further, optical spectra lack emission such as [\feii] $\sim$\lb4350 emission,  seen in nebular spectra of SNe Ia of \cite{Taubenberger2013} and \cite{Black2016}, for example. Other researchers have also had difficulty modeling the \feii\ feature near 4350 \AA\ 
\citep[Spyromillo 2016, private communication; Sim 2016, private communication; ][]{Mazzali2015,Friesen2017}. It is not surprising that these models struggle to get the ionization correct -- there are no free parameters and the density structure and element distribution is set by the adopted initial model.

At late times the super-\Mch\ model was generally in better agreement with observation -- a result of the model being cooler with lower ionization. Since we know that most of the observed SNe we discussed are not super-\Mch, there is a fundamental problem with the models. This problem might arise from the adopted explosion models, be related to assumptions about mixing and clumping, and/or be a problem in the ionization calculations. Since nebular spectra show strong [\feii] and [\feiii], the Fe$^+$/Fe$^{2+}$ ratio must be of order unity, and consequently it is sensitive to the Fe atomic models (and the density structure). 

The disconnect between early and late time modeling is not unexpected. Early time spectra are dependent on the outer ejecta whereas late time spectra are primarily dependent on the inner ejecta. Further, the processes determining the observed spectra  in the photospheric and nebular phases are distinct, and subject to different uncertainties in the atomic data.

Another problem is the strong nebular [\siii] \lb\lb9068,9530 and [\ariii] \lb\lb7135,7751 lines. The [\siii] \lb\lb9068,9530 does not seem to appear in nebular spectra. However, it is not clear if [\ariii] \lb7135 is present. There are three additional transitions contributing to that overall feature between 7000-7500 \AA. There are two [\feii] \lb\lb7155,7172 lines that overlap [\ariii] \lb7135 and, depending on the ionization structure of the ejecta, it becomes difficult to determine the source of the feature in observations. However, atomic physics of the [\ariii] \lb\lb7135,7751 lines requires that the line ratio, $I(7135)/I(7751)$, should be a factor of 4.2, so if spectral detections of [\ariii] \lb7751 are possible, then one can determine the strength of the blended [\ariii] \lb7135 line. However, observed SN Ia nebular spectra appear absent of IME lines. This could be due to an absence of \elem[56]{Ni} in the IME zone. In our models, the presence of  some \elem[56]{Ni} in the IME zone means that positrons are available as a heating source after the ejecta has become optically thin to \gray\ photons. One would expect some level of mixing to occur through Rayleigh-Taylor instabilities between these layers -- see \cite{Hicks2015} and references therein. To address the problem of too high an ionization, clumping, arising from radiation hydrodynamic instabilities, should be considered in future studies. Our preliminary work shows that, as expected, clumping lowers the ionization, and we will address this issue in a future paper.

Another possible explanation concerns the validity of the explosion models. We have considered only four models, and only two of the explosion models were obtained from ``first principles", and even these were derived from 1D explosions. Alternative explosion mechanisms might give rise to different density and abundance profiles, and in particular, the spatial distribution of \elem[56]{Ni}. The later will influence the amount of UV line blanketing, potentially introducing degeneracies with the ejecta mass. However, other diagnostics (e.g. the NIR nickel line) provide additional information, and can break the degeneracies. Further, despite the deficiencies, the models have highlighted important diagnostics and questions that can help facilitate future progress towards understanding Type Ia SNe.

% In reality, density and abundance profiles may be  different. We have limited ourselves to just four models and have not modeled the influence of varying spatial distributions of IGEs like \elem[56]{Ni}. The distribution of IGEs by mixing processes might have a strong impact on UV blanketing, potentially introducing degeneracies with the ejecta mass. This is a limiting factor of this research.}

%
\section{Conclusion}\label{Conclusion}%%%%%%%%%%%%%%%%%%%%%%%%%%%%%%%%%%%%%%%%%%%%%%%%
We have presented four 1D SN Ia models -- three delayed detonation models with masses of  1.02, 1.40, and 1.70 \Msun\ and one detonation sub-\Mch\ model with a mass of 1.04 \Msun. By design, the models have the same \elem[56]{Ni} mass of $\sim$0.62 \Msun\ which allows us to investigate the dependence of light curves and spectra on ejecta mass. Despite the smallness of  the model grid they serve to highlight important diagnostics that can help facilitate future progress towards understanding Type Ia SNe.

Our results show that despite large differences in ejecta mass, the optical flux throughout the photospheric phase shows less than 0.3 mag difference in peak brightness in the LCs, as well as nearly identical spectral features. We have seen that the peak bolometric luminosity of each model is similar to within about 15 per cent, and the difference in rise time is less than $\sim$20 per cent. Due to differences in diffusion time, however, the two sub-\Mch\ mass models do evolve faster (pre-maximum) by a day as seen from the bolometric luminosity and synthetic $B$-band LCs. There is only a slight difference ($\sim$5 per cent) in the decline parameter, $\Delta M_{15}(B)$, between sub-\Mch\ and super-\Mch\ models. Our sub-\Mch\ models have much bluer colours at all epochs compared with SUP ($B-R$ difference of $\approx 0.3$ mag at maximum and a difference in $B-R\gtrsim$1 mag roughly 20 days post maximum). Our models show larger differences in NIR light curves, particularly with the $H$-band's $\sim$1 mag difference at maximum light between sub-\Mch\ and super-\Mch.

Spectroscopically,  at most photospheric phases, the optical spectra show the same gross features. However, the strength of UV blanketing between 2000-4000 \AA\ is found to correlate with ejecta mass. Lower mass models have higher temperature and ionization (as more heating per gram), and hence lower UV blanketing between 2000-4000 \AA. $M(^{56}{\rm Ni})/M_{\rm ej}$ is the leading parameter controlling this study. Higher mass models produce stronger IME features, such as the \caii\ NIR triplet and the \skii\ \lb6347,6371 doublet prior to the nebular phase, and stronger [\sii] \lb\lb9530,9068, [\caii] \lb\lb7291,7324, and [\ariii] \lb7135 in the nebular phase. Lower mass models have higher ionization, as indicated by the presence of the \skiii\ \lb\lb4553,4568,4575 triplet near maximum and the lack of strong \feii\ and \coii\ lines in the optical post-photospheric/nebular phase. Model SUB1, unlike SUB2, is dominated by strong [\feiii] and [\coiii] lines, such as [\feiii] \lb4658, [\feiii] \lb5270, [\coiii] \lb5888, and [\coiii] 1.5484 $\mu$m. 

In the nebular phase, the  [\nkii] 1.939 $\mu$m line is absent in our sub-\Mch\ detonation model, but readily visible in the three other models. Potentially, the [\nkii] 1.939 $\mu$m line provides us with a diagnostic of the amount of stable nickel (\elem[58]{Ni} \& \elem[60]{Ni}), unlike the blended optical [\nkii] \lb\lb7378,7412 lines. Its absence in NIR spectra would provide strong evidence for a lack of a `\elem[56]{Ni} hole' and potentially sub-Chandrasekhar mass ejecta (given 1D modeling). However, complex ionization issues can influence the strength of all [\nkii] lines, making absolute determinations of the abundance model-dependent. Overall the NIR provides the best diagnostics for distinguishing between our different SN Ia progenitor models. 

In comparing our spectra to observation at times greater than 20 days post maximum, we consistently find better qualitative fits with our cooler, high mass super-\Mch\ model. Given that there is a $\sim$20-70 per cent difference in claimed ejecta mass between our compared observational objects and our super-\Mch\ model, we suggest clumping as a way to lower the high ionization and high temperatures observed in our models. 
While it is difficult to reproduce all observational features due to the diversity of SNe Ia, we are able to match some features shown in our comparison to observations. Prior to maximum, the best choice of model varies. Agreement depends on the velocity structure of the ejecta. For instance, the photospheric features \skii\ \lb\lb6347,6371 and the \caii\ NIR triplet expose the difficulty of reproducing the velocity structure of SN Ia (\figs~\ref{SN2001ay_spec} and \ref{SN2005el_spec}, for example). Future efforts to reproduce the diversity of these features requires a better understanding of the outer ejecta and explosion mechanism.

Parallel work has been undertaken by Blondin et al. (submitted to MNRAS) who studied SN 1999by using a low mass  model (0.9\,\Msun) and a model with a  Chandrasekhar mass (both with 0.12\,\Msun of \isoni). They find that the lower mass model provides a better match to the light curve, and exhibits a faster rise and a brighter maximum. As in our study, the lower mass model does not show the [\nkii] 1.939 $\mu$m line, which is seen in their Chandrasekhar model

To determine more accurate diagnostic signatures of SN Ia progenitors we need to understand clumping and inhomogeneities in Ia ejecta. Some insights can be obtained from multi-dimensional explosion modeling, while additional insight might be obtained from studies of young SN remnants that are not interacting with the surrounding ISM.  During the photospheric phase, more UV spectral data will help to constrain the ionization and temperature of the gas. More NIR spectral data will help to test our diagnostics, such as the NIR \caii\ triplet (or nebular [\caii] \lb\lb7291,7324), the nebular features between 9000 \AA-1 $\mu$m (such as [\siii] \lb\lb9068,9530), and the [\nkii] 1.939 $\mu$m line (requiring SNe Ia at a high enough redshift to avoid the telluric absorption). These nebular features can provide leverage on the progenitor channel by constraining initial densities ([\nkii]), the overlap between IMEs and IGEs ([\caii]), and the ionization structure. 

As many more SN spectra become available it will be possible to do systematic statistical comparisons between SNe which have a similar initial \elem[56]{Ni} mass. As discussed above, our studies show that Type Ia SN will exhibit systematic differences in spectra and multi-band LCs as a function of ejecta mass, thus providing fundamental constraints on the nature of the progenitors.

\section*{acknowledgements}%%%%%%%%%%%%%%%%%%%%%%%%%%%%%%%%%%%%%%%%%%
We thank St\'{e}phane Blondin for providing feedback for this work as well as everyone at the ``Supernovae Through the Ages: Understanding the Past to Prepare for the Future" conference who presented work and talked with us about our research. We also thank the referee for providing very detailed and useful comments and questions. This research has made use of the CfA Supernova Archive, which is funded in part by the National Science Foundation through grant AST 0907903.  DJH acknowledges partial support from STScI theory grant HST-AR-12640.01, and DJH and KDW thank NASA for partial support through theory grant NNX14AB41G. 
\bibliography{SNeIa_diag_TEST}

\label{lastpage}

\end{document}